\documentstyle[12pt,epsfig,rotating]{article}                                                   
\newcommand{\figfI}     {and1.epsi}                 % Z wavefunction renorm
\newcommand{\figfII}    {and2.epsi}                 % MSSM fermion selfenergy
\newcommand{\figfIII}   {and3.epsi}                 % MSSM Z->ff vertex corr.
\newcommand{\figI}      {scanmu+.eps}               % charginos vs. mu
\newcommand{\fighI}     {scanmu_high.eps}           % charginos vs. mu
\newcommand{\figII}     {rb_vgl.eps}                % rb contourplots
\newcommand{\fighII}    {rb_vgl_hi.eps}             % rb contourplots
\newcommand{\figIV}     {mssmtb1_11_sw.eps}         % balkendiagramm tanb=1.0
\newcommand{\figV}      {mssmbsg1.6_11_sw_znn.eps}  % balkendiagramm 11
\newcommand{\figVI}     {mssmbsg50_11_sw.eps}       % balkendiagramm 11 tanb=50
\newcommand{\figchiI}   {chi2_214.eps}              % chi**2 
\newcommand{\figchiII}  {chi2_215.eps}              % chi**2 
\newcommand{\figchiIII} {chi2_216.eps}              % chi**2 
\newcommand{\figchiIV}  {chi2_217.eps}              % chi**2 
\newcommand{\mtmh}      {mtmh.eps}                  % SM Contourplot
\newcommand{\smdchi}    {smdchi.eps}                % SM Deltachi(m_H)
\newcommand{\sintw}     {sintw.eps}                 % SM sin^2theta(mH,mt)
\newcommand{\figochiall}{chi2_25.eps}               % all optimized fit
\newcommand{\figochi}   {chi2_26.eps}               % optimized fit
\newcommand{\figorb}    {Rb_26.eps}                 % optimized fit
\newcommand{\figobsg}   {bsg_26.eps}                % optimized fit
\newcommand{\figoalf}   {alphas_26.eps}             % optimized fit
\newcommand{\figomix}   {phimix_26.eps}             % optimized fit
\newcommand{\fighochi}  {chi2_21_hi.eps}            % optimized fit
\newcommand{\fighorb}   {Rb_21_hi.eps}              % optimized fit
\newcommand{\fighobsg}  {bsg_21_hi.eps}             % optimized fit
\newcommand{\fighoalf}  {alphas_21_hi.eps}          % optimized fit
\newcommand{\fighomix}  {phimix_21_hi.eps}          % optimized fit
\newcommand{\alphas}  {alphas.eps}                   % measurements of alphas
\catcode`\@=11                                                                  
\long\def\@makefntext#1{                                                        
\protect\noindent \hbox to 3.2pt {\hskip-.9pt                                   
$^{{\ninerm\@thefnmark}}$\hfil}#1\hfill}                %CAN BE USED                          
                                                                                
\def\@makefnmark{\hbox to 0pt{$^{\@thefnmark}$\hss}}  %ORIGINAL                 
                                                                                       
\def\ps@myheadings{\let\@mkboth\@gobbletwo                                      
\def\@oddhead{\hbox{}                                                           
\rightmark\hfil\ninerm\thepage}                                                 
\def\@oddfoot{}\def\@evenhead{\ninerm\thepage\hfil                              
\leftmark\hbox{}}\def\@evenfoot{}                                               
\def\sectionmark##1{}\def\subsectionmark##1{}}                                  
                                                                                
\newcounter{sectionc}\newcounter{subsectionc}\newcounter{subsubsectionc}        
\renewcommand{\section}[1] {\vspace*{0.6cm}\addtocounter{sectionc}{1}           
\setcounter{subsectionc}{0}\setcounter{subsubsectionc}{0}\noindent              
        {\normalsize\bf\thesectionc. #1}\par\vspace*{0.4cm}}                           
\renewcommand{\subsection}[1] {\vspace*{0.6cm}\addtocounter{subsectionc}{1}     
        \setcounter{subsubsectionc}{0}\noindent                                        
        {\normalsize\it\thesectionc.\thesubsectionc. #1}\par\vspace*{0.4cm}}           
\renewcommand{\subsubsection}[1]                                                
{\vspace*{0.6cm}\addtocounter{subsubsectionc}{1}                                
        \noindent {\normalsize\rm\thesectionc.\thesubsectionc.\thesubsubsectionc.      
        #1}\par\vspace*{0.4cm}}

\newcounter{appendixc}                                                          
\newcounter{subappendixc}[appendixc]                                            
\newcounter{subsubappendixc}[subappendixc]

\renewcommand{\appendix}[1] {\vspace*{0.6cm}                                    
        \refstepcounter{appendixc}                                              
        \setcounter{figure}{0}                                                  
        \setcounter{table}{0}                                                   
        \setcounter{equation}{0}                                                
        \renewcommand{\thefigure}{\Alph{appendixc}.\arabic{figure}}             
        \renewcommand{\thetable}{\Alph{appendixc}.\arabic{table}}               
        \renewcommand{\theappendixc}{\Alph{appendixc}}                          
        \renewcommand{\theequation}{\Alph{appendixc}.\arabic{equation}}         
        \noindent{\bf Appendix \theappendixc #1}\par\vspace*{0.4cm}}

\renewenvironment{thebibliography}[1]                                           
        {\begin{list}{\arabic{enumi}.}                                                 
        {\usecounter{enumi}\setlength{\parsep}{0pt}                                    
\setlength{\leftmargin 1.25cm}{\rightmargin 0pt}                                
         \setlength{\itemsep}{0pt} \settowidth                                         
        {\labelwidth}{#1.}\sloppy}}{\end{list}}                                        
                                                                                
\newcounter{itemlistc}                                                          
\newcounter{romanlistc}                                                         
\newcounter{alphlistc}                                                          
\newcounter{arabiclistc}

\newcommand{\fcaption}[1]{                                                      
        \refstepcounter{figure}                                                 
        \setbox\@tempboxa = \hbox{\footnotesize Fig.~\thefigure. #1}            
        \ifdim \wd\@tempboxa > 6in                                              
           {\begin{center}                                                      
        \parbox{6in}{\footnotesize\baselineskip=12pt Fig.~\thefigure. #1}       
            \end{center}}                                                       
        \else                                                                   
             {\begin{center}                                                    
             {\footnotesize Fig.~\thefigure. #1}                                
              \end{center}}                                                     
        \fi}                                                                    
                                                                                
\newcommand{\tcaption}[1]{                                                      
        \refstepcounter{table}                                                  
        \setbox\@tempboxa = \hbox{\footnotesize Table~\thetable. #1}            
        \ifdim \wd\@tempboxa > 6in                                              
           {\begin{center}                                                      
        \parbox{6in}{\footnotesize\baselineskip=12pt Table~\thetable. #1}       
            \end{center}}                                                       
        \else                                                                   
             {\begin{center}                                                    
             {\footnotesize Table~\thetable. #1}                                
              \end{center}}                                                     
        \fi}                                                                    
                                                                                
 1                                          
 1                                          
 1

\font\ninerm=cmr9

\begin{document}
\sloppy

\begin{titlepage}
\begin{flushright}
%\vspace*{-2.2cm}
\noindent
 \hfill IEKP-KA/96-07    \\
 \hfill KA-TP-18-96\\
 \hfill hep-ph/9607286   \\
 \hfill July 7th, 1996 \\
\end{flushright}
\vspace{1.7cm}
\begin{center} {\Large\bf Global Fits of the SM and MSSM \\
         to Electroweak Precision Data \\}
\vspace{0.5cm}
{\small {\bf  W.~de Boer$^1$ \footnote{Email: wim.de.boer@cern.ch}, 
 A.~Dabelstein$^3$\footnote{E-mail: Andreas.Dabelstein@tu-muenchen.de}, 
 W.~Hollik$^2$\footnote{E-mail: hollik@itpaxp3.physik.uni-karlsruhe.de}, \\
 W.~M\"osle$^2$\footnote{E-mail: wm@itpaxp1.physik.uni-karlsruhe.de}, 
 U.~Schwickerath$^1$\footnote{E-mail: Ulrich.Schwickerath@cern.ch},
\\
}
{\it 1) Inst.\ f\"ur Experimentelle Kernphysik, Univ.\ of Karlsruhe, \\}
{\it Postfach 6980, D-76128 Karlsruhe, Germany  \\} 
{\it 2) Inst.\ f\"ur Theoretische Physik, Univ.\ of Karlsruhe, \\}
{\it Postfach 6980, D-76128 Karlsruhe, Germany  \\}} 
{\it 3) Inst.\ f\"ur Theoretische Physik, Univ.\ of M\"unchen, \\}
{\it James Franck Stra\ss e, D-85747 Garching, Germany}
\end{center}

\vspace{1.5cm}

\begin{center}
{\bf Abstract}\\
\end{center}

%Present LEP data show a too high value of $R_b$ (3.4$\sigma$) and a too low
%value of $R_c$ (2$\sigma$) where $R_{b(c)}$ is the ratio
%$R_{b(c)}=\Gamma_{Z^0\rightarrow b\bar b (c\bar c)}/\Gamma_{Z^0\rightarrow q\bar q}$.

The Minimal supersymmetric extension of the Standard Model (MSSM) with light stops,
charginos or pseudoscalar Higgs bosons has been
suggested as an explanation of the too high value of the 
branching ratio of the $Z^0$ boson into $b$~quarks ($R_b$ anomaly).
A program including all radiative corrections to the MSSM at the same
level as the radiative corrections to the SM has been developed and used
to perform global fits to all electroweak data from LEP, SLC and the Tevatron.
%If one excludes the $b\rightarrow s\gamma$ rate from CLEO in the fit,
The probability of the global fit improves from $8\%$ in the SM to $18\%$
in the MSSM.
%However, such light charginos and stops influence strongly the  $b\rightarrow s\gamma$ rate
%as measured by CLEO.
Including the $b\rightarrow s\gamma$ rate, as measured by CLEO, reduces the probability
from $18\%$ to $15\%$.
%We find that the predicted masses of charginos and/or stops are not necessarily in the
%LEP II range.
In the constrained MSSM
requiring unification and electroweak
symmetry breaking no improvement of $R_b$ is possible.

\parindent0.0pt
\newpage
\thispagestyle{empty}
\vspace{5cm}
\newpage

\end{titlepage}

\section{Introduction}
 Present LEP data show a too high value of $R_b$ (3.4$\sigma$) and a too low
value of $R_c$ (2$\sigma$) where $R_{b(c)}$ is the ratio
$R_{b(c)}=\Gamma_{Z^0\rightarrow b\bar b (c\bar c)}/\Gamma_{Z^0\rightarrow q\bar q}$.
In the past it has been shown by several groups that it is possible to improve $R_b$
using supersymmetric models with light charginos, stops or Higgses, which yield
positive contributions to the $Zb\bar b$ vertex
%\cite{yel1,boufi,chan2,ell1,garc3,kan1,kan2,garcia,garcia2}.
\cite{yel1}\nocite{boufi,chan2,ell1,garc3,kan1,kan2,garcia}-\cite{garcia2}.
In this paper we perform an equivalent analysis of all electroweak data
both in the Standard Model (SM) and its supersymmetric extension (MSSM).
The conclusion is that only a moderate increase in the probability can
be found for the MSSM as compared to the SM, if all present limits on supersymmetric particles and the
ratio $b\rightarrow s\gamma$ are taken into account.
The analysis was performed using all actual electroweak data from Tevatron \cite{tevatron,top},
LEP and SLC \cite{Lep1,Lep2},
the measurement of
$\frac{BR(b\rightarrow s\gamma)}{BR(b\rightarrow ce\bar\nu)}$ from CLEO \cite{cleo}
and limits on the masses of supersymmetric
particles \cite{yel1,lim1,lim2,lim4,lim6,rev96}.

For low values of $\tan\beta$ the diagrams with charginos and right handed
stops in the vertex loop are dominant, while for high $\tan\beta$ the exchange
of the pseudoscalar Higgs between the outgoing $b$-quarks, which is proportional to
$m_b\cdot\tan\beta$ becomes important too. It should be noted that light stops and charginos
will contribute to the $b\rightarrow s\gamma$ rate with an opposite sign as the SM
diagram of the top - W boson loop, so the predicted $b\rightarrow s\gamma$ rate can
become easily too small, if the stops and charginos are light. Therefore, getting
$R_b$ and  $b\rightarrow s\gamma$  right is not easy,
since $R_b$ wants sparticle masses near the experimental limits, while
$b\rightarrow s\gamma$ needs sparticles well above the experimental limits or a sufficiently large
$\tan\beta$.

\section{$Z^0$ boson on-resonance observables}

At the $Z$ boson resonance two classes of precision observables are available:
\begin{itemize}
\item[a)] inclusive quantities:
  \begin{itemize}
   \item[$\bullet$] the partial leptonic and hadronic decay width $\Gamma_{f 
                  \bar{f}}$,
   \item[$\bullet$] the total decay width $\Gamma_Z$,        
   \item[$\bullet$] the hadronic peak cross section $\sigma_h$, 
   \item[$\bullet$] the ratio of the hadronic to the electronic decay
                  width of the $Z$ boson: $R_h$, 
   \item[$\bullet$] the ratio of the partial decay width for $Z\rightarrow c\bar{c} \,
                  ( b \bar{b} )$ to the hadronic width, 
                  $R_c$, $R_b$.
  \end{itemize}
\item[b)] asymmetries and the corresponding mixing angles:
  \begin{itemize}
   \item[$\bullet$] the \it forward-backward \rm asymmetries $A_{FB}^f$,
   \item[$\bullet$] the \it left-right \rm  asymmetries $A_{LR}^f$,
   \item[$\bullet$] the $\tau$ polarization $P_\tau$,
   \item[$\bullet$] the effective weak mixing angles $\sin^2 \theta_{eff}^f$.
  \end{itemize}
\end{itemize}
Together with the quantity $\Delta r$ in the correlation of the $W$ mass to the 
electroweak input parameters $G_\mu$, $M_Z$ and $\alpha_{em}$, this set of
precision observables is convenient for a numerical analysis of the 
supersymmetric parameter space.
In the following the observables defined above are expressed
with the help  of effective couplings.

\subsection{The effective $Z$-$f$-$f$ couplings}                                     

The coupling of the $Z$ boson to fermions $f$ can be expressed by effective
vector and axial vector coupling constants $v_{eff}^f, \, a_{eff}^f$ in terms of the
NC vertex:
\begin{equation}
J_{NC}^\mu = \frac{e}{2 s_W^2 c_W^2} \, \gamma^\mu \, ( v_{eff}^f - a_{eff}^f \gamma_5 ) \ ,
\end{equation}
where the convention is introduced : $c_W^2 = \cos^2 \theta_W = 1 - s_W^2 = M_W^2 / 
M_Z^2$ \cite{sirlin}.
Input parameters are the $\mu$ decay constant $G_\mu = 1.166392 \times 10^{-5}$
GeV$^{-2}$, $\alpha_{EM} = 1/137.036$ and the mass of the $Z^0$ boson $M_Z = 91.1884$
GeV. The mass of the $W$ boson is related to these input parameters through:
\begin{eqnarray}\label{gmudr}                                                   
{G_{\mu}\over\sqrt{2}} = {\pi\alpha_{EM}\over 2 s^2_W M^2_W} \cdot
\frac{1}{
1 - \Delta r_{MSSM} \left(\alpha_{EM},M_W,M_Z,m_t,...\right)} \ ,
\end{eqnarray} 
where the complete MSSM one-loop contributions are parameterized by the
quantity $\Delta r_{MSSM}$\cite{sola}.
Leading higher order Standard Model corrections\cite{bhp,fleischer} to the 
quantity $\Delta r$ are included in the calculation.
\smallskip 

The effective couplings $v_{eff}^f, \, a_{eff}^f$ can be written as:
\begin{eqnarray}
v_{eff}^f & = & \sqrt{Z_Z} \, (v^f + \Delta v^f + Z_M Q_f) \nonumber \\
a_{eff}^f & = & \sqrt{Z_Z} \, (a^f + \Delta a^f) \ .
\end{eqnarray}
$v^f$ and $a^f$ are the tree-level vector and axial vector couplings:
\begin{equation}
v^f = I_3^f - 2 Q_f s_W^2 \ , \ a^f =   I_3^f.
\end{equation}
$Z_Z$, $Z_M$ are given in eq. (\ref{ZZ}).
The complete MSSM one-loop contributions of the non-universal finite vector and 
axial vector couplings $\Delta v^f$, $\Delta a^f$ have been calculated\cite{dab/hollik},
together with the leading two-loop Standard Model 
contributions\cite{bhp,fleischer,Chet/Kw}.
They are derived in the 't Hooft-Feynman gauge and in the
on-shell renormalization scheme\cite{bhs}.
Fig.~ 1 shows the MSSM one-loop $Z \rightarrow f \bar{f}$ vertex correction diagrams.\\

\vspace{2cm}
%\begin{figure}[ht]
%\epsfig{figure=zbb1.ps,%
%height=7.0cm,width=11.0cm,%
%bbllx=0pt,bblly=520pt,bburx=405pt,bbury=751pt}
%\end{figure}

%\vspace{-2.0cm}

%\begin{figure}[ht]
%\epsfig{figure=zbb2.ps,%
%height=7.0cm,width=11.0cm,%
%bbllx=0pt,bblly=520pt,bburx=405pt,bbury=751pt}
%\caption{MSSM one-loop $Z \rightarrow f \bar{f}$ vertex correction diagrams. $i,j,k =
% 1,..,2 (4)$ are chargino, neutralino and sfermions indices. No particle permutations
% are shown.}
%\end{figure}
\begin{figure}[ht]
% \protect\vspace{-1cm}
 \begin{center}
  \leavevmode
  \epsfxsize=16cm
  \epsffile{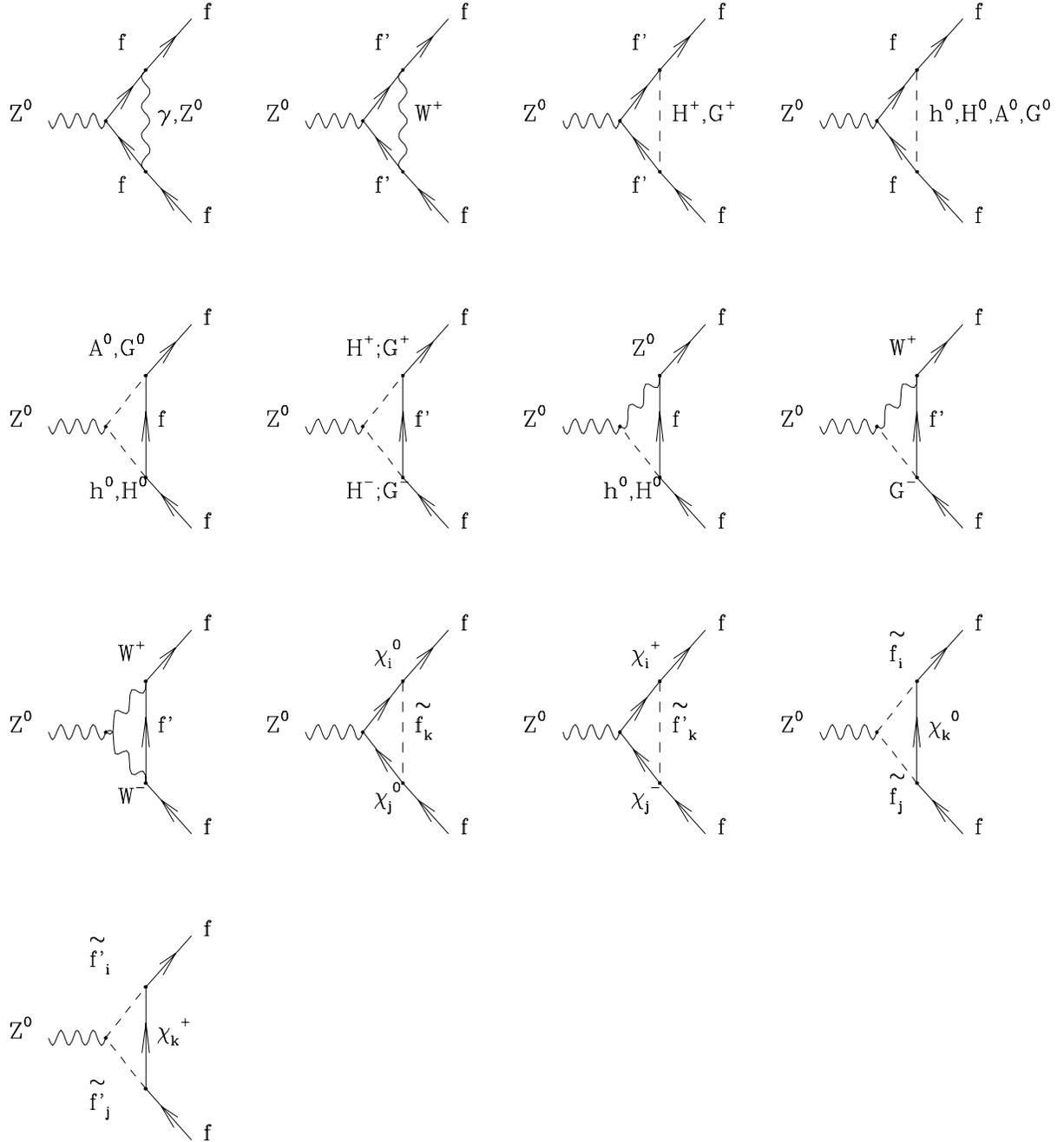}
\end{center}
%\vspace{-8cm}
\caption{\label{\figfIII} MSSM one-loop $Z \rightarrow f \bar{f}$ vertex correction diagrams. $i,j,k =
 1,..,2 (4)$ are chargino, neutralino and sfermions indices. No particle permutations
 are shown.
}
\end{figure}

\begin{figure}
% \protect\vspace{-1cm}
 \begin{center}
  \leavevmode
  \epsfxsize=16cm
  \epsffile{\figfII}
\end{center}
%\protect\vspace{-16cm}
\caption{\label{\figfII}
MSSM  fermion self-energies.}
%\protect\vspace{-4cm}
\begin{center}
  \leavevmode
  \epsfxsize=16cm
  \epsffile{\figfI}
\end{center}
%\protect\vspace{-12cm}
\caption{\label{\figfI}$Z$ boson wave function renormalization.}
\end{figure}

\vspace{1.0cm} 
\clearpage
\noindent
The non-universal contributions  can be written  in the following way:
\begin{eqnarray}
 \Delta v_f & = & F_V^{SM} + \Delta F_V , \nonumber \\
 \Delta a_f & = & F_A^{SM} + \Delta F_A . \nonumber
\end{eqnarray}
The Standard Model form factors $F_{V,A}^{SM}$ corresponding to the
diagrams of figs. 1 and 2 can be found
e.g in refs.~\cite{bhs,bhp}.
The diagrams with a virtual photon are listed for completeness in the figures.
They are not part of the effective weak couplings but are treated separately
in the QED corrections, together with real photon bremsstrahlung.
The non-standard contributions are summarized by
\begin{eqnarray}
\Delta F_V & = &  \sum_i F_V^{(i)} 
                + v_f \delta Z_V^f + a_f \delta Z_A^f \nonumber \\
\Delta F_A & = &  \sum_i F_A^{(i)} 
                + v_f \delta Z_A^f + a_f \delta Z_V^f \nonumber 
\end{eqnarray}
where the sum extends over the diagrams of fig. 1 with internal
charged and neutral Higgs bosons, charginos, neutralinos and scalar 
fermions, each diagram contributing 
$$
    F_V^{(i)} \gamma_{\mu} - F_A^{(i)} \gamma_{\mu}\gamma_5
$$
to the $Zff$-vertex.
The self-energy diagrams of fig. 2 with internal neutral Higgs,
chargino, neutralino and sfermion lines determine the field renormalization
constants
\begin{eqnarray}
 \delta Z_V^f & = & -\Sigma_V(m_f^2) -
                  2m_f^2 [ \Sigma'_V(m_f^2) + \Sigma'_s(m_f^2) ]
                  \nonumber \\
 \delta Z_A^f & = & \Sigma(m_f^2) 
\end{eqnarray}
with the scalar functions $\Sigma_{V,A,S}$ in the decomposition of the 
fermion self-energy according to
 $$
  \Sigma = \not{p} \Sigma_V + \not{p} \gamma_5 \Sigma_A + m_f \Sigma_S . $$
The contributions from the Higgs sector are given explicitly in 
ref.~\cite{dghk}. For the genuine SUSY diagrams, the
couplings for  charginos, neutralinos and sfermions are taken 
from \cite{hunter}, together with the diagonalization matrices given in 
section 3.

\medskip 
%\newpage
The universal propagator corrections from the 
 finite $Z$ boson wave function renormalization $Z_Z$ and the $\gamma Z$
mixing $Z_M$ are derived from the $(\gamma, \, Z)$ propagator matrix. The 
inverse matrix is:
\begin{equation}
(\Delta_{\mu \nu})^{-1} = i g_{\mu \nu} \, \left(
 \begin{array}{ll}
 k^2 + \hat{\Sigma}_\gamma (k^2) &  \hat{\Sigma}_{\gamma Z} (k^2) \\
 \hat{\Sigma}_{\gamma Z} (k^2)   &  k^2 - M_Z^2 + \hat{\Sigma}_Z (k^2)
 \end{array} \right) \ ,
\label{gzmatrix}
\end{equation}
where $\hat{\Sigma}_\gamma$, $\hat{\Sigma}_{Z}$, $\hat{\Sigma}_{\gamma Z}$
are the renormalized self energies and mixing.
They are obtained by summing the loop diagrams, shown symbolically in fig.\ref{\figfI}, and the counter terms
and can be found in ref.~\cite{dabelstein}.
\smallskip

The entries in the $(\gamma, \, Z)$ propagator matrix:
\begin{equation}
\Delta_{\mu \nu} = - i g_{\mu \nu} \left( \begin{array}{ll}
\Delta_\gamma     &  \Delta_{\gamma Z} \\
\Delta_{\gamma Z} &  \Delta_Z \end{array} \right) \ ,
\end{equation}
are given by:
\begin{eqnarray}
\Delta_\gamma (k^2) & = & \frac{1}{k^2 +  \hat{ \Sigma}_\gamma (k^2) -
 \frac{\hat{ \Sigma}_{\gamma Z}^2 (k^2)}{k^2 - M_Z^2 + \hat{ \Sigma}_Z (k^2)} }
 \nonumber \\
 \Delta_Z (k^2) & = & \frac{1}{k^2 - M_Z^2 + \hat{ \Sigma}_Z (k^2) -
 \frac{\hat{ \Sigma}_{\gamma Z}^2 (k^2)}{k^2  + \hat{ \Sigma}_\gamma (k^2)} }
 \nonumber \\
\Delta_{\gamma Z} (k^2) & = & -\frac{\hat{ \Sigma}_{\gamma Z} (k^2)}{
 [ k^2  + \hat{ \Sigma}_\gamma (k^2)] \, 
 [ k^2 - M_Z^2 + \hat{ \Sigma}_Z (k^2) ] - \hat{ \Sigma}_{\gamma Z}^2 (k^2) } \ .
\label{zprop}
\end{eqnarray}
The renormalization condition to define the mass of the $Z$ boson is given
by the pole of the propagator matrix (eq.~\ref{gzmatrix}). 
The pole $k^2 = M_Z^2$ is the solution of the equation: 
\begin{equation}
 \cal R\rm e \, [ \, ( M_Z^2   + \hat{ \Sigma}_\gamma (M_Z^2) \, ) \, 
 \hat{ \Sigma}_Z (k^2)  - \hat{ \Sigma}_{\gamma Z}^2 (M_Z^2) \, ] = 0 \ .
\end{equation}
Eq. (\ref{zprop}) yields the wave function renormalization $Z_Z$ and mixing $Z_M$:
\begin{eqnarray}
Z_Z & = & Res_{M_Z} \Delta_Z = \left. \frac{1}{1 +
        \cal R\rm e \hat{ \Sigma}_Z' (k^2) -
        \cal R\rm e
   \left( \frac{ \hat{ \Sigma}_{\gamma Z}^2 (k^2)}{k^2 + \hat{ \Sigma}_\gamma (k^2)
   } \right)'  } \ \right| _{k^2 = M_Z^2} \nonumber \\
Z_M & = & - \frac{\hat{ \Sigma}_{\gamma Z} (M_Z^2)}{M_Z^2 + 
 \hat{ \Sigma}_\gamma (M_Z^2) } \ .
\label{ZZ}
\end{eqnarray}

\subsection{ $Z$ boson observables}               

\noindent
The fermionic $Z$ boson partial decay widths $\Gamma_{f \bar{f}}$ can be 
written as follows:
\begin{itemize}
\item[1)] $f \ne b$:

\begin{eqnarray}
 \Gamma_{f \bar{f}} & = & \frac{N_C \, 
 \sqrt{2} G_{\mu} M_Z^3}{12\pi} (1-\Delta r_{MSSM}) \,
 \left[ (v_{eff}^f)^2 
  + (a_{eff}^f)^2 (1 - \frac{6 m_f^2}{M_Z^2} ) \right] \cdot \nonumber \\
  & & \cdot (1 + \frac{3 \alpha_{EM}}{4 \pi} Q_f^2) \, (1 + \delta_{QCD}^f ) \ ,
\end{eqnarray}
where 
\begin{equation}
 \delta_{QCD}^f = \left\{ \begin{array}{ll}
  0 & ,f = \mbox{leptons} \nonumber \\
  \frac{\alpha_s}{\pi} + 1.405 (\frac{\alpha_s}{\pi})^2 - 12.8
  (\frac{\alpha_s}{\pi})^3
   -\frac{Q_f^2}{4}\frac{\alpha\alpha_s}{\pi^2} & ,f = \mbox{quarks}
 \end{array} \right. \ .
\label{deqc}
\end{equation}
\item[2)] $f = b$:
\begin{eqnarray}
 \Gamma_{b \bar{b}} & = & \frac{N_C \, 
 \sqrt{2} G_{\mu} M_Z^3}{12\pi} (1-\Delta r_{MSSM}) \,
 \left[ (v_{eff}^b)^2 + (a_{eff}^b)^2  \right] \nonumber \\
   &  & 
 \cdot (1 + \frac{3 \alpha_{EM}}{4 \pi} Q_b^2) \, (1 + \delta_{QCD}^b ) \,
 + \Delta \Gamma_{b \bar{b}} \ . \nonumber \\
\end{eqnarray}
In $\Delta \Gamma_{b \bar{b}}$ the $b$ quark specific finite mass terms 
with QCD corrections \cite{Chet/Kw} are included.
 $\delta_{QCD}^b$ is given in eq. (\ref{deqc}).
\end{itemize}
The total decay width $\Gamma_Z$ is the sum of the contributions from leptons and quarks: \\
 \begin{equation}
   \Gamma_Z = \sum_{f} \Gamma_{f \bar{f}} \ .
 \end{equation}
 In the following $\Gamma_{had} = \sum_{q} \Gamma_{q \bar{q}}$ is the hadronic decay
 width of the $Z$ boson.
\medskip

\noindent
The hadronic peak cross section is defined as
\begin{equation}
 \sigma_h = \frac{12 \pi}{M_Z^2} \frac{\Gamma_{ee} \Gamma_{had}}{\Gamma_Z^2} \ .
\end{equation}
The ratio of the hadronic to the electronic decay width is defined as
\begin{equation}
 R_e = \frac{\Gamma_{had}}{\Gamma_{e e}} \ .
\end{equation}
The ratio of the partial decay width for
$Z \rightarrow b \bar{b} \, (c \bar{c})$ to
the total hadronic decay width is given by
\begin{equation}
 R_{b (c)} = \frac{\Gamma_{b \bar{b} (c \bar{c}) }}{\Gamma_{had}} \ .
\end{equation}
\medskip

The following quantities and observables depend on the ratio of the vector to
axial vector coupling. The effective flavour dependent weak mixing angle can
be written as
\begin{equation}
 \sin^2 \theta_{eff}^f = \frac{1}{4 | Q_f |} \, \left( 1 - \frac{v_{eff}^f}{a_{eff}^f} 
 \right) \ .
\end{equation}
The \it left-right \rm asymmetries are given by

\begin{equation}
 A_{LR}^f = {\cal A}^{\rm f}  = \frac{2 \, v_{eff}^f / a_{eff}^f}{1 + ( v_{eff}^f /
 a_{eff}^f )^2 } \ ,
\end{equation}
while the \it forward-backward \rm asymmetries can be written as
\begin{equation}
 A_{FB}^f = \frac{3}{4} \, \cal A^{\rm e} \rm \, \cal A^{\rm f} \rm \ .
\end{equation}

\section{The MSSM}

%In the following we discuss the general version of the MSSM without further
%restrictions to the parameters from assumptions about Grand Unification or Super-gravity.
\subsection{Higgs sector}
The scalar sector of the MSSM is completely determined by the value
of $\tan\beta=v_2/v_1$ and the pseudoscalar mass $M_A$, together with the 
radiative corrections. The latter ones are taken into account in terms of
the effective potential approximation with the leading terms $\sim m_t^4$,
including the mixing in the scalar top system \cite{ellisetal}.
In this way, the coupling constants of the various Higgs particles
to gauge bosons and fermions can be taken over from \cite{hunter}
substituting only the scalar mixing angle $\alpha$ by the improved 
effective mixing angle which is obtained from the diagonalization of the
scalar mass matrix.

\subsection{Sfermion sector}

%Sfermions and charginos/neutralinos appear in the vertex correction to
%$Z$ boson decays and give sizeable contributions for the external $b$ state.
The physical masses of squarks and sleptons are described by a $2 \times 2$ mass matrix:
\begin{equation}
 \cal M_{\rm \tilde{f}}^{\rm 2} \rm = \left( \begin{array}{ll}
 M_{\tilde{Q}}^2 + m_f^2 + M_Z^2 (I_3^f - Q_f s_W^2) \cos 2 \beta &
 m_f (A_f + \mu \{ \cot \beta , \tan \beta \} ) \\
 m_f (A_f + \mu \{ \cot \beta , \tan \beta \} ) &
 M_{\{\tilde{U},\tilde{D}\}}^2 + m_f^2 + M_Z^2 Q_f s_W^2 \cos 2 \beta
 \end{array} \right) \ ,
\label{sqmatrix}
\end{equation}
with SUSY soft breaking parameters $M_{\tilde{Q}}$,
$M_{\tilde{U}}$, $M_{\tilde{D}}$, $A_f$, and $\mu$.
It is convenient to use  the following notation for the 
off-diagonal entries in eq. (\ref{sqmatrix}):
\begin{equation}
A_f' = A_f + \mu \{ \cot \beta , \tan \beta \} \ .
\label{glaprime}
\end{equation}
Scalar neutrinos appear only as left-handed mass eigenstates.
Up and down type sfermions in (\ref{sqmatrix}) are distinguished by
setting f=u,d and the $\{u,d\}$ entries in the parenthesis.
Since the non-diagonal terms are proportional to $m_f$, it seems
natural to assume unmixed sfermions for the lepton and quark case
except for the scalar top sector.
The $\tilde{t}$ mass matrix is diagonalized by a rotation matrix with
a mixing angle $\Phi_{mix}$. 
Instead of $M_{\tilde{Q}}$, $M_{\tilde{U}}$, $M_{\tilde{D}}$,  $A_t'$
for the $\tilde{b}$, $\tilde{t}$ system
the physical squark masses
$m_{\tilde{b}_L}, m_{\tilde{b}_R}$, $m_{\tilde{t}_2}$ can be used
together with 
$A_t'$ or, alternatively, the stop mixing angle $\Phi_{mix}$.
For simplicity we assume $m_{\tilde{b}_L}=m_{\tilde{b}_R}$, and 
$\tilde{u}$, $\tilde{d}$, $\tilde{c}$, $\tilde{s}$
to have masses  masses  equal to the 
$\tilde{b}$ squark mass.

\smallskip
A possible mass splitting between
 $\tilde{b}_L$-$\tilde{t}_L$  yields a contribution to the $\rho$-parameter
$\rho = 1+ \Delta\rho$
in terms of \cite{sola}: \footnote{The superscript
$\Delta \rho^0$ indicates that no left-right mixing is present.}
\begin{equation} 
\Delta \rho_{\tilde{b}-\tilde{t}}^0 = \frac{3 \alpha_{EM}}{16 \pi s_W^2 M_W^2} \,
 \left(m_{\tilde{b}_L}^2 + m_{\tilde{t}_L}^2 - 2 \frac{m_{\tilde{b}_L}^2 
 m_{\tilde{t}_L}^2}{m_{\tilde{b}_L}^2 - m_{\tilde{t}_L}^2} \log \frac{
 m_{\tilde{b}_L}^2}{m_{\tilde{t}_L}^2} \right) \ . 
\end{equation}
As a universal loop contribution, it enters the quantity
\begin{equation} 
 \Delta r \simeq \Delta \alpha_{EM} -
  \frac{c_W^2}{s_W^2} \, \Delta \rho \, + ... \nonumber    
\end{equation}
and all the $Z$ boson widths 
$$  \Gamma_{f\bar{f}} \sim 1 + \Delta\rho + \cdots $$
and is thus significantly constrained by the data on $M_W$ and the
leptonic widths.
%The choice $m_{\tilde{b}_L} \simeq m_{\tilde{t}_L}$ and small stop
%mixing avoids 
%large contributions to the $\rho$-parameter.

\subsection{Chargino/Neutralino sector}
\noindent
The chargino (neutralino) masses and the mixing angles in the gaugino
couplings are calculated from soft breaking parameters $M_1$, $M_2$ and
 $\mu$ in
the chargino (neutralino) mass matrix\cite{hunter}. 
The validity of the GUT relation $M_1 = 5/3 \tan^2 \theta_W \, M_2$ is assumed. 

The chargino $2 \times 2$ mass matrix is given by
\begin{equation}\label{charmat}
  \cal{M}_{\rm \tilde{\chi}^\pm} \rm = \left( \begin{array}{ll}
    M_2 & M_W \sqrt{2} \sin \beta \\ M_W \sqrt{2} \cos \beta & - \mu \\
    \end{array}  \right) \ ,
\end{equation} \\
with the SUSY soft breaking parameters $\mu$ and $M_2$ in the diagonal
matrix elements. The physical chargino mass states $\tilde{\chi}^{\pm}_i$ 
are the rotated
wino and charged Higgsino states:
\begin{eqnarray}
\tilde{\chi}^+_i & = & V_{ij} \psi^+_j   \nonumber \\
\tilde{\chi}^-_i & = & U_{ij} \psi^-_j  \ ; \ i,j = 1,2  \ .
\end{eqnarray}
$V_{ij}$ and $U_{ij}$ are unitary chargino mixing matrices obtained from
the diagonalization of the mass matrix eq.~\ref{charmat}:
\begin{equation}
\rm U^* \cal{M}_{\rm \tilde{\chi}^\pm} \rm  V^{-1} = diag(m_{\tilde{\chi}^\pm_1},m_{\tilde{\chi}^\pm_2}) \ .
\end{equation}
\smallskip \par
The neutralino $4 \times 4 $ mass matrix can be written as:   
\begin{equation}\label{neutmat}
 \cal{M}_{\rm \tilde{\chi}^0}   = \left( \begin{array}{cccc}
 M_1 & 0 & - M_Z \sin \theta_W \cos \beta & M_Z \sin \theta_W \sin \beta \\
 0 & M_2 & M_Z \cos \theta_W \cos \beta & - M_Z \cos \theta_W \sin \beta \\
- M_Z \sin \theta_W \cos \beta & M_Z \cos \theta_W \cos \beta & 0 &  \mu \\
 M_Z \sin \theta_W \sin \beta & - M_Z \cos \theta_W \sin \beta & \mu & 0
 \\  \end{array} \right) 
\end{equation}
where the diagonalization can be obtained by the unitary matrix $N_{ij}$: 
\begin{equation}
  \rm N^* \cal{M}_{\rm \tilde{\chi}^0} \rm N^{-1}  = diag(
  m_{\tilde{\chi}_i^0}) \ .
\end{equation}

\smallskip \noindent
The elements $U_{ij}$, $V_{ij}$, $N_{ij}$ of the diagonalization
matrices  enter the couplings of the 
charginos, neutralinos and sfermions to fermions and gauge bosons, as
explicitly given in ref.~\cite{hunter}. Note that our sign convention on the
parameter $\mu$ is opposite to that of ref.~\cite{hunter}.

\section{Results}

\subsection{Chargino Masses}

\begin{figure*}
\thispagestyle{empty}
  \protect\vspace{-1cm}
 \begin{center}
  \leavevmode
  \epsfxsize=9.0cm
  \epsffile{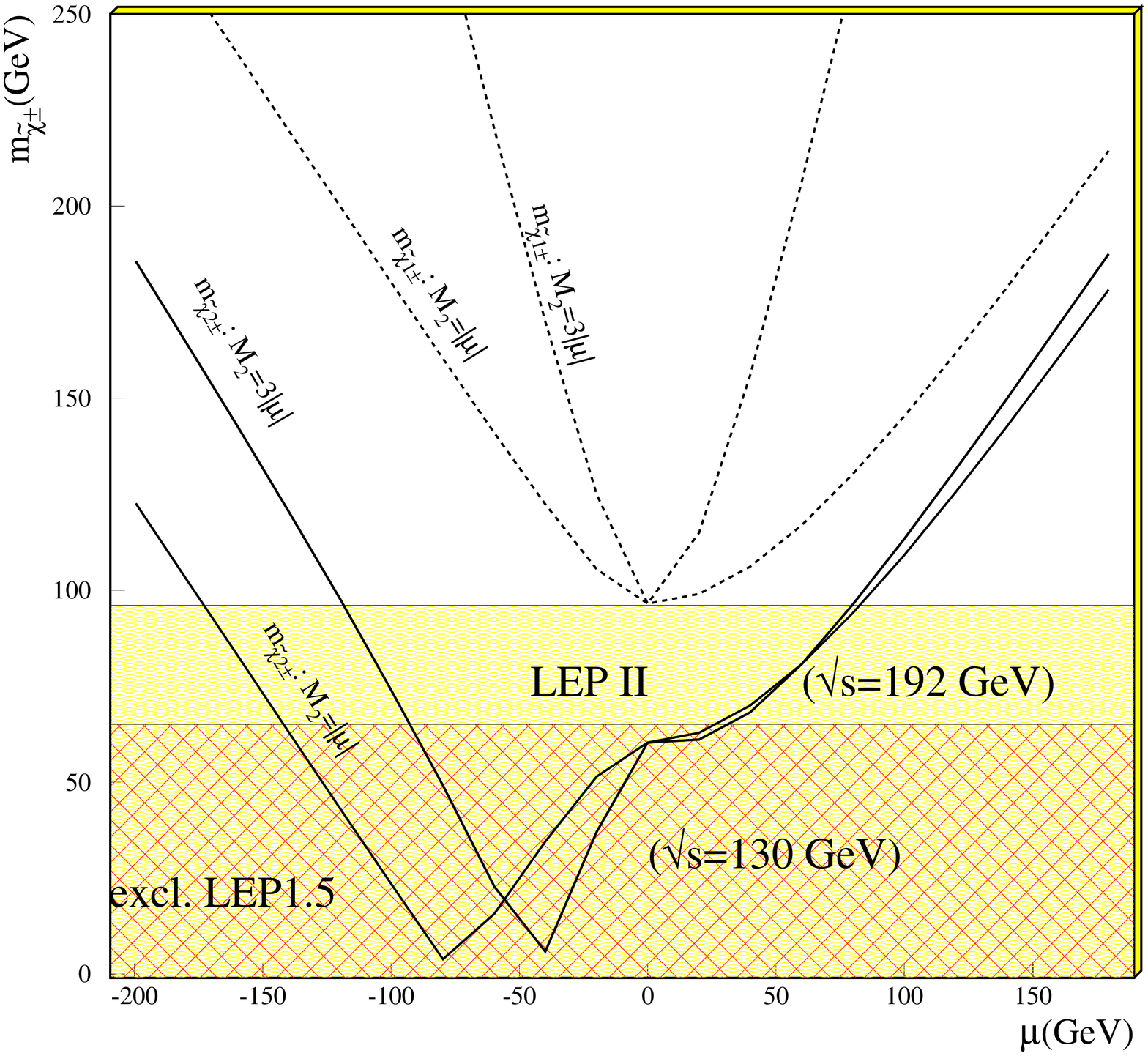}
\end{center}
\protect\vspace{-1cm}
\caption{\label{\figI} 
Dependence of the chargino masses on the parameter
$\mu$ for $M_2=\mid\mu\mid$ and $M_2=3\mid\mu\mid$ 
for $\tan\beta=1.6$, $\alpha_s\approx 0.117$ and $\tilde{m}_{t_2}\approx 60~$GeV. No
optimization of parameters was performed here. The
shaded regions indicate chargino masses less than $65$~GeV which are
excluded by $LEP~1.5$ and chargino masses less than $96$~GeV.
It can be observed that for positive values of $\mu$
two light
charginos are easier to obtain, if $\mu$ and $M_2$ have similar values. }
\begin{center}
  \leavevmode
  \epsfxsize=9.0cm
  \epsffile{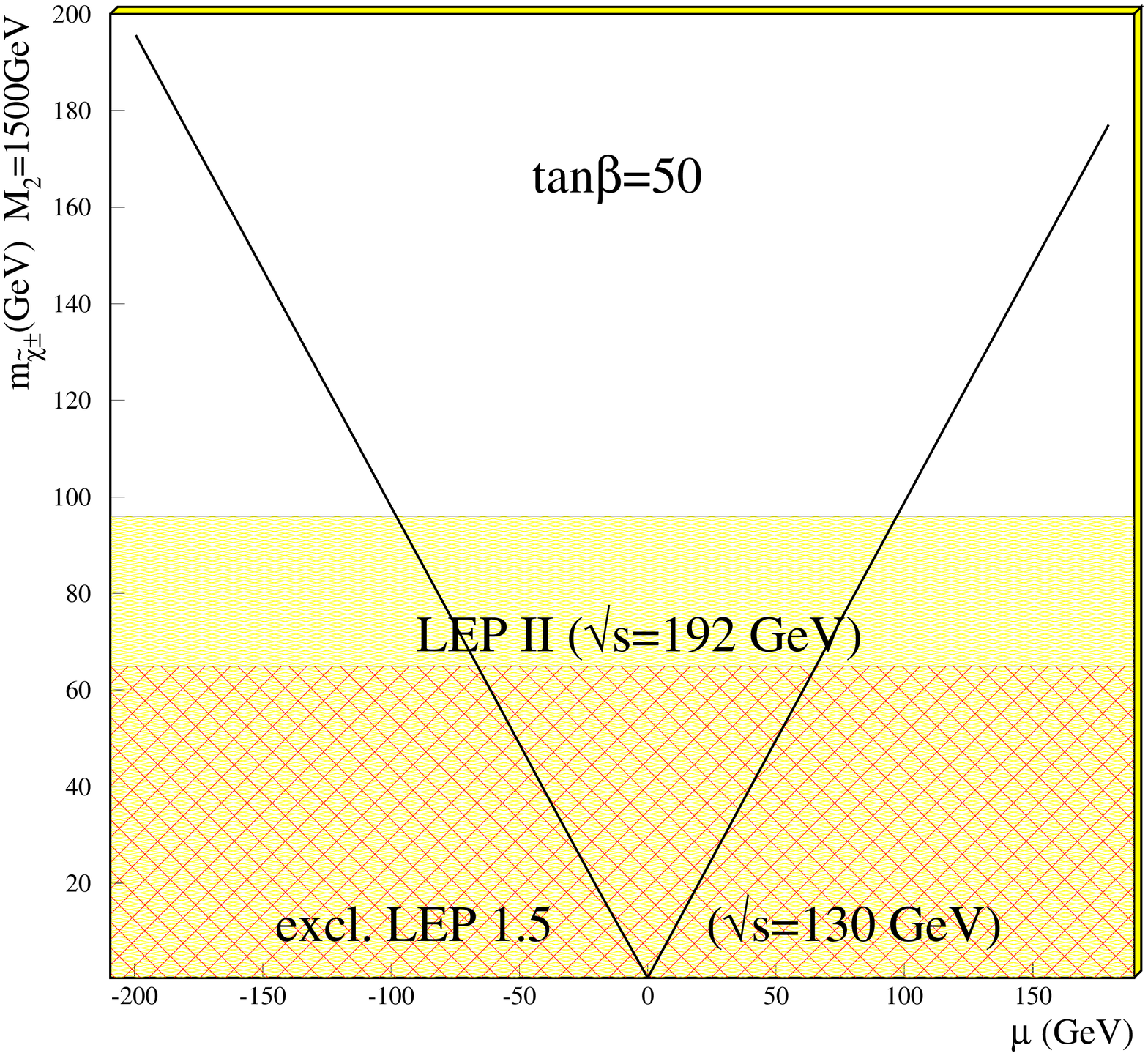}
\end{center}
\protect\vspace{-1cm}
\caption{\label{\fighI} 
Dependence of the chargino masses on
the parameter $\mu$ for $\tan\beta=50$. In this case $M_2$=1500~GeV
was used and no
optimization of parameters was performed. The
shaded regions indicate chargino masses less than $65$~GeV which are
excluded by $LEP~1.5$ and chargino masses less than $96$~GeV.
The light chargino mass depends on $\mu$, whereas the heavy chargino
mass is dominated by the value of $M_2$ and keeps approximately constant
at 1500~GeV. For a fixed given chargino mass there are two possible
solutions corresponding to \protect{$\mu < 0 $} and $\mu > 0$, respectively.}
\end{figure*}

\begin{figure*}
 \begin{center}
  \leavevmode
  \epsfxsize=10cm
  \epsffile{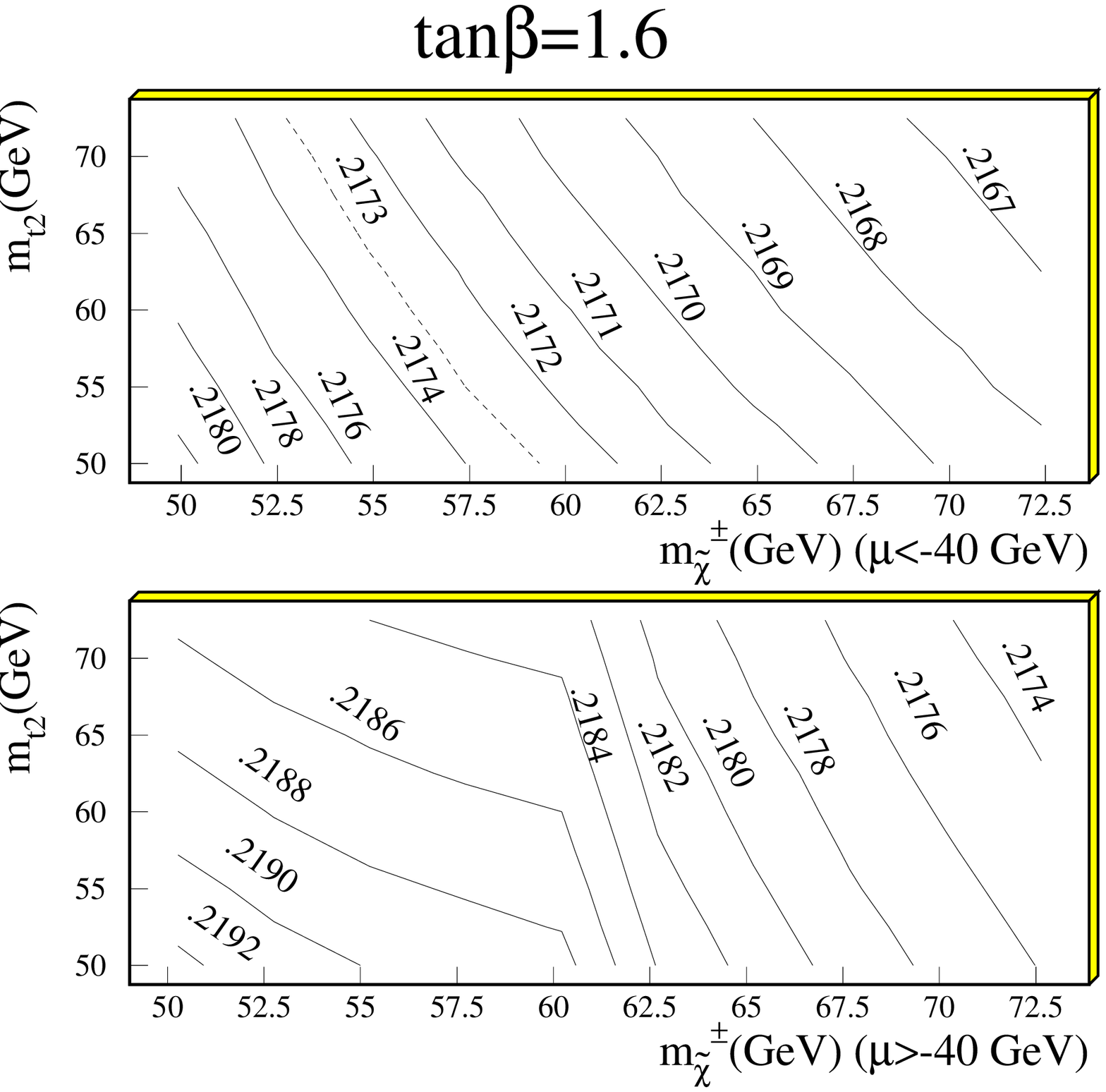}
 \end{center}
\caption{\label{\figII} $R_b$ in the light stop versus light chargino plane with
 $M_2=3\mid\mu\mid$ and $\tan\beta=1.6$.
The upper part shows the solution with $\mu<-40~$GeV, in the
lower part the one with $\mu>-40~$GeV
is displayed. In the latter solution
quite high values for $R_b$ are possible, as can be seen in the figure.
The dashed line in the upper plot indicates the $2\sigma$ lower limit of $R_b$.
}
 \begin{center}
  \leavevmode
  \epsfxsize=10cm
  \epsffile{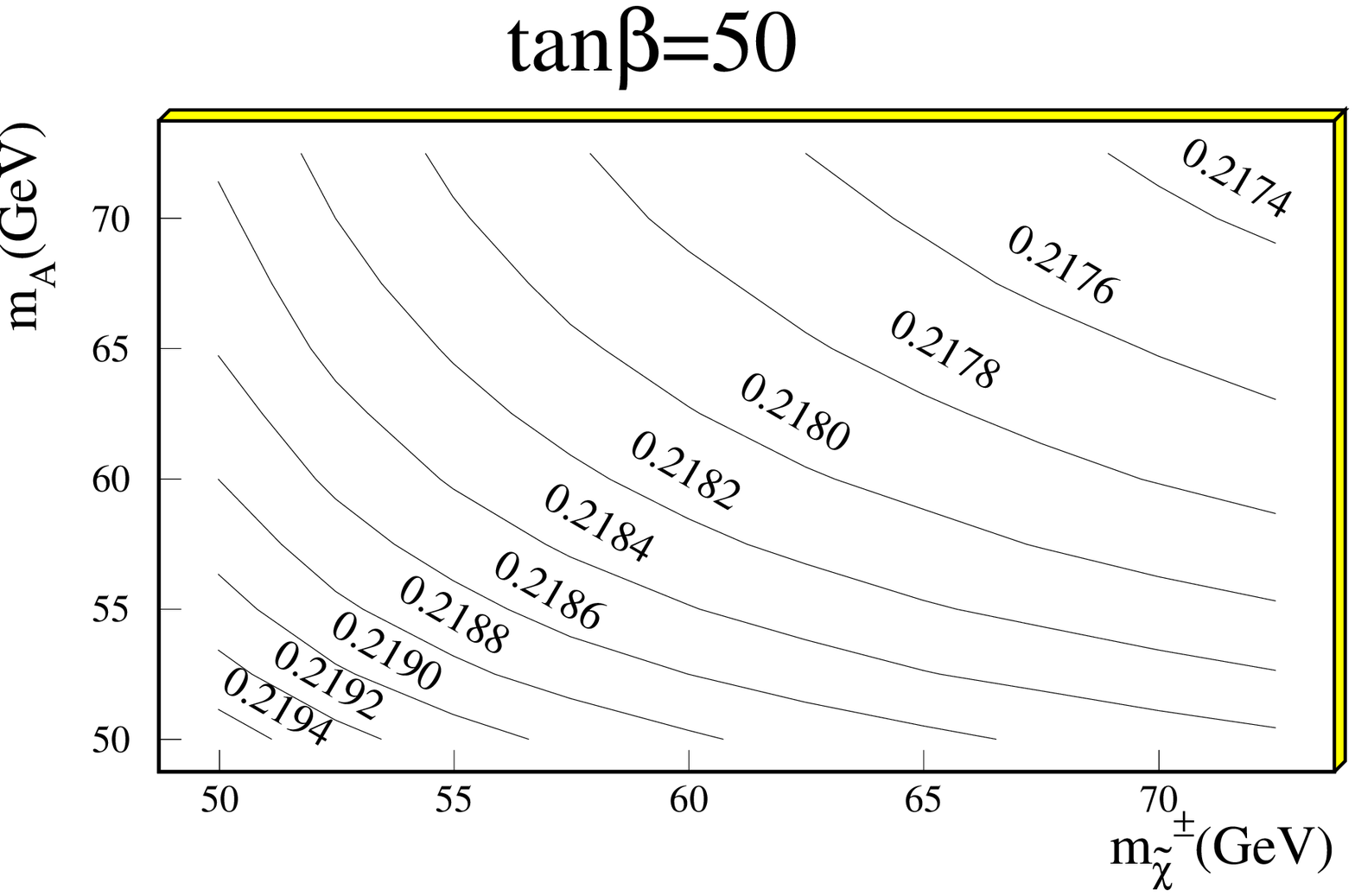}
 \end{center}
\caption{\label{\fighII} $R_b$ in the $m_A$ versus light chargino plane with
  $M_2=3\mid\mu\mid$ for the high $\tan\beta$ solution. $\mu$ was chosen positive
  here. In this case choosing the opposite sign for $\mu$ doesn't change $R_b$.
} 
\end{figure*}

As mentioned before the low $\tan\beta$ scenario  of the MSSM
needs a light right handed stop and light higgsino-like chargino for
a large value of $R_b$, whereas in
the high $\tan\beta$ scenario one needs in addition a light pseudoscalar Higgs $A$
\cite{yel1,chan2}.
A higgsino - like chargino can be obtained for a low value of the parameter $\mu$ in the
mass matrix (eq. \ref{charmat}).
Figs.~\ref{\figI}
and ~\ref{\fighI} show the dependence of the chargino masses on the parameter $\mu$ in a
region of the parameter space which yields a good global $\chi^2$.
In case of high $\tan\beta$, 
$m_{\tilde\chi_{2}}$ is
almost symmetric around zero, whereas in case of low $\tan\beta$
this dependence is more complicated, as can be seen from fig.~\ref{\figI}.
For $M_2=3\mid\mu\mid$ the light chargino mass passes zero  at
$\mu=-40$, so the following low $\tan\beta$ plots were made for
$\mu>-40$ and $\mu\le-40$ GeV.
The asymmetric structure of fig.~\ref{\figI} is reflected in
the contours of constant $R_b$ in the $m_{\tilde\chi_{2}}$ versus light scalar
top $m_{\tilde{t_2}}$ plane (see fig.~\ref{\figII}).
High values of
$R_b$ up to 0.2194 are possible (see figs.~\ref{\figII} and \ref{\fighII}),
although these special regions of the parameter space are already
experimentally excluded by the lower limits on sparticle masses.

\begin{figure*}
\begin{minipage}[t]{8cm}
\leavevmode
\epsfxsize = 7cm
\epsffile{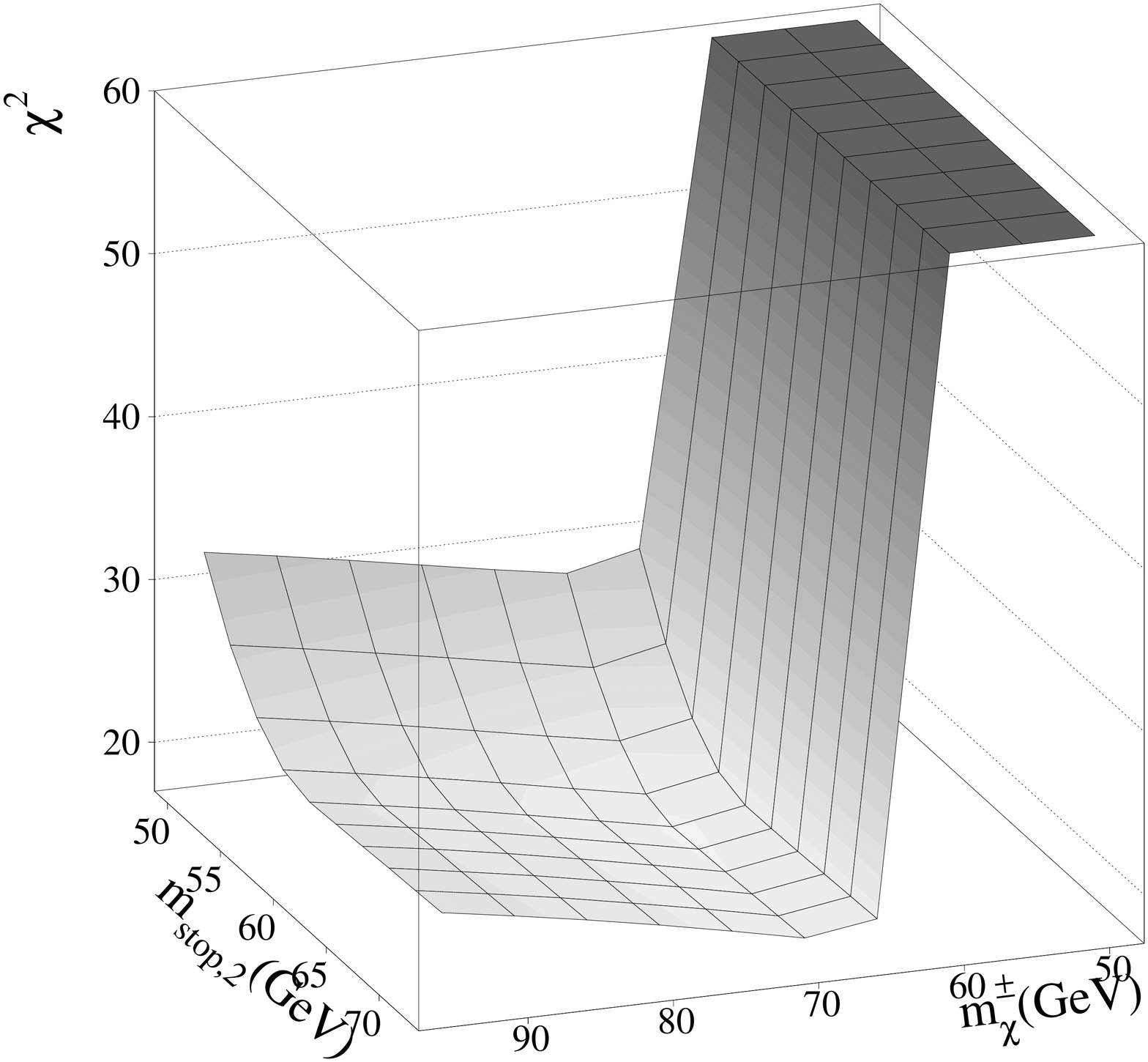}
\end{minipage}
\hfill
\begin{minipage}[t]{8cm}
\epsfxsize = 7cm
\epsffile{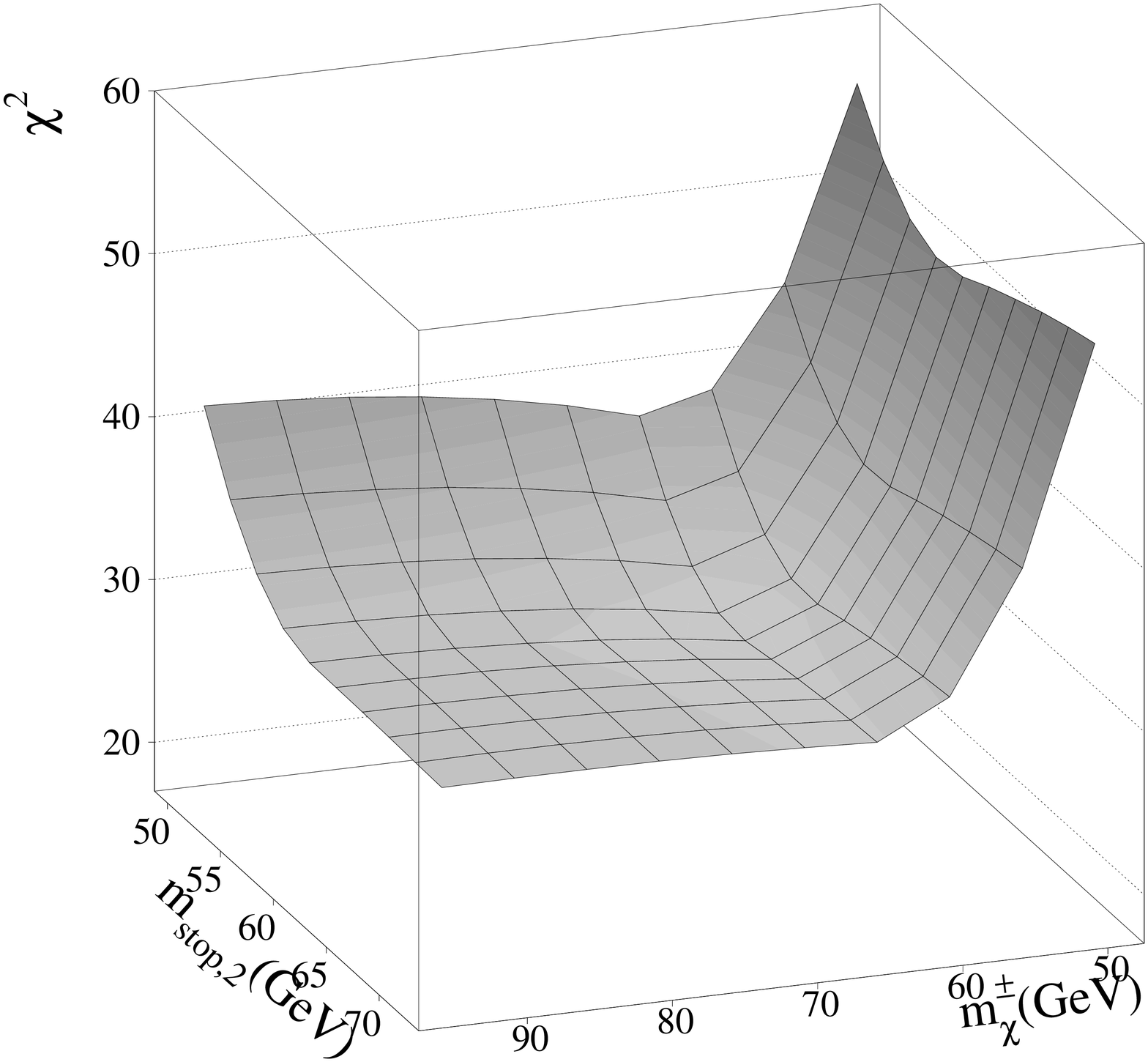}
\end{minipage}
\caption{\label{\figchiI} Dependence of the absolute $\chi^2$ for $\mu>-40~$GeV
(left side) and  $\mu<-40~$GeV (right side), using $M_2=3\mid\mu\mid$.
%All parameters were fixed to values which gave a good $\chi^2$.
At the different
points of the grid no optimisation of parameters was performed, but they were fixed to
values near a good minimum.}
\end{figure*}

\begin{figure*}
\begin{minipage}[b]{8cm}
\epsfxsize = 7cm
\epsffile{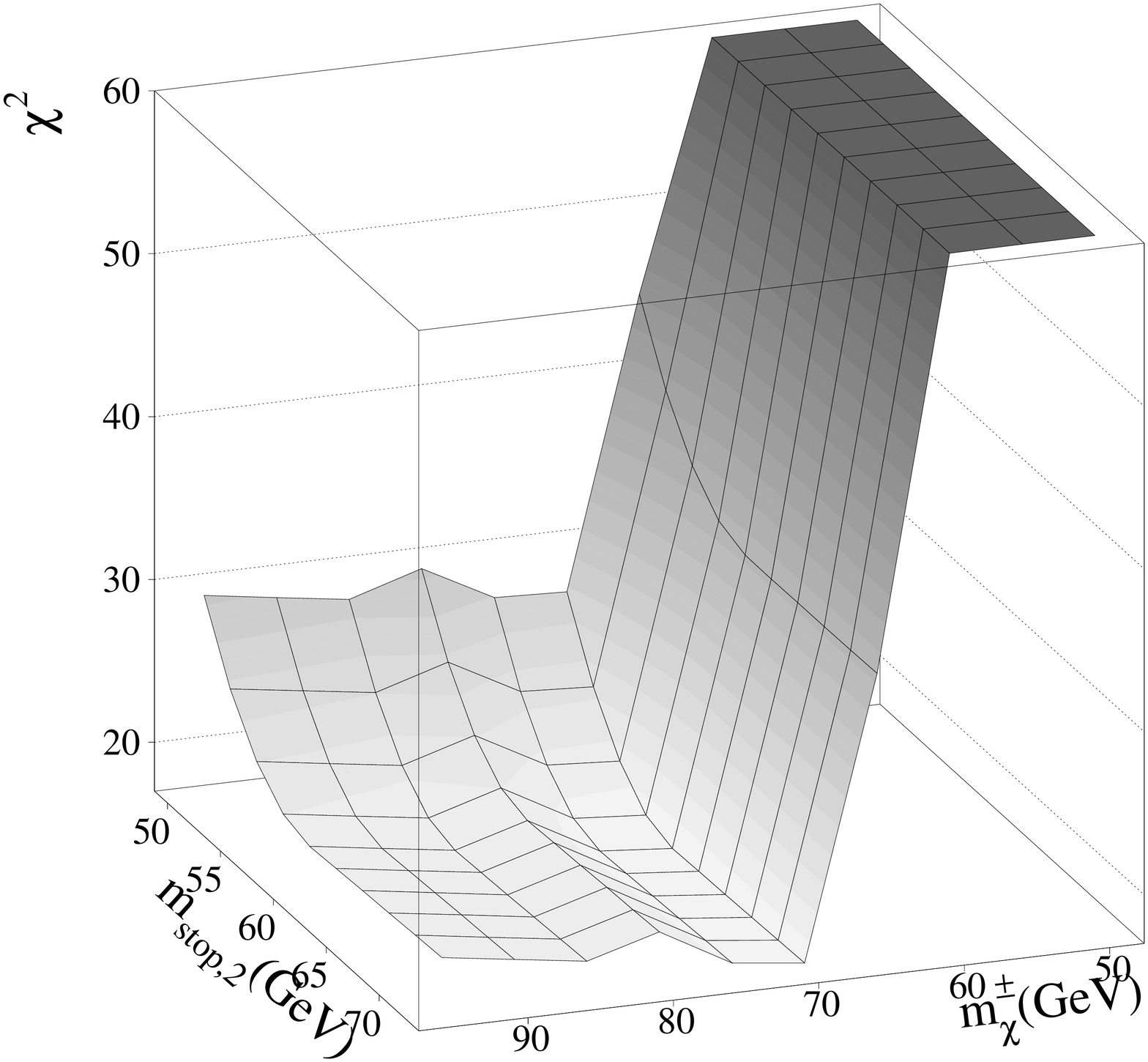}
\end{minipage}
\hfill
\begin{minipage}[b]{8cm}
\epsfxsize = 7cm
\epsffile{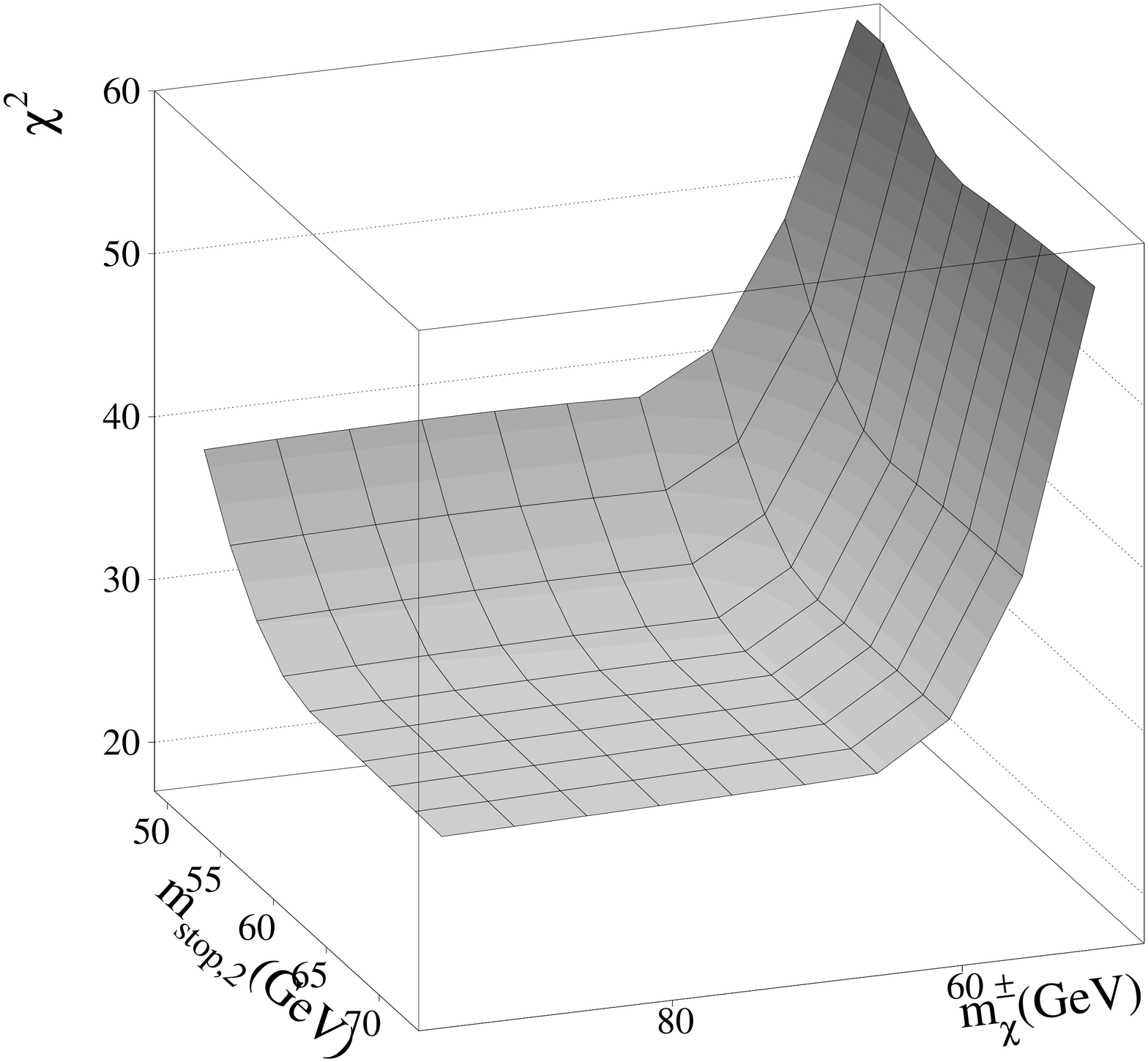}
\end{minipage}
\caption{\label{\figchiIII} The same as \protect\ref{\figchiI}, 
but for $M_2=\mid\mu\mid$.
}
\end{figure*}

Taking
$M_2=\mid\mu\mid$ does not change these results very much, as can be seen from
a comparison of the $\chi^2$ distributions in
fig.~\ref{\figchiI} ($M_2=3\mid\mu\mid$) and fig.~\ref{\figchiIII} ($M_2=\mid\mu\mid$).
%Here the overall $\chi^2$ without optimization of parameters is plotted for both
%solutions of $\mu$ in the case of low $\tan\beta$.
The small increase of the $\chi^2$ at
chargino masses around 80~GeV in the left hand part of fig.~\ref{\figchiIII} is due to
neutralino threshold singularities, for which an additional $\chi^2$  contribution
has been added,
if the sum of two neutralino masses is close to the $Z^0$ mass.
The sharp increase of the $\chi^2$ function at low chargino masses is due to experimental
limits on chargino, neutralino and stop 
masses from LEP 1.5~\cite{lim1,lim2,lim4}.

\subsection{Optimization of Parameters}
An optimization of free parameters of the MSSM was performed by minimizing a
$\chi^2$ function using MINUIT \cite{minuit}. Several contributions to the $\chi^2$ were
taken into account:
\begin{itemize}
\item experimental limits on the masses of supersymmetric particles and neutralino production
      from LEP 1.5 and Tevatron \cite{lim1,lim2,lim4}
\item precision measurements of on resonance observables from LEP \cite{Lep2}, taking error
      correlations into account
\item the measurement of the branching ratio $\frac{BR(b\rightarrow s\gamma)}{BR(b\rightarrow ce\bar\nu)}$ from CLEO \cite{cleo}
\end{itemize}

\begin{table}[ht]
\begin{center}
\begin{tabular}{|c|c|}
\hline
\hline
\multicolumn{2}{|c|}{experimental limits}\\   
\hline
\hline
$\tilde m_{\chi^{\pm}_{1,2}}$ & $>$ 65~GeV \\
\hline
$\tilde m_{\chi^{0}_1}$ & $>$ 13~GeV \\
$\tilde m_{\chi^{0}_2}$ & $>$ 35~GeV \\
$\tilde m_{\chi^{0}_{3,4}}$ & $>$ 60~GeV \\
%$\tilde m_{\chi^{0}_4}$ & 60~GeV \\
$\Gamma_{Z\rightarrow neutralinos}$ & $<2$~MeV\\
\hline
$\tilde m_{t_{1,2}}$ & $>$ 48~GeV \\
\hline
$m_h$,$m_H$,$m_A$,$m_{H^\pm}$  & $>$ 50~GeV\\
\hline
\end{tabular}
\end{center}
\caption{\label{limits}
Mass limits assumed for the optimized fits.}
\end{table}

The experimental limits included in the fit are summarized in table \ref{limits}
\cite{yel1,lim1,lim2,lim4,rev96}.
%Stop masses just above 45~GeV are not excluded
%provided that the lightest neutralino is heavy enough \cite{lim4}. 
The calculation of the total decay width of the Z boson into neutralinos is based on
reference \cite{zwid1}, the calculation of the ratio $b\rightarrow s\gamma$ on reference
\cite{bsgamma}.

As already mentioned, $R_b$ mainly depends on the stop mass and the light chargino mass
for the low $\tan\beta$ and on the pseudoscalar Higgs mass and chargino mass for
the high $\tan\beta$ scenario. In order to get a feeling for the size of the effects, we
study the dependence of the
$\chi^2$ function on these parameters with optimization of the remaining parameters.
This has been done for both the low and the high $\tan\beta$ solutions.
%In the next section the best solutions will be presented.
The best solutions will be presented in the next chapter.\\

\begin{figure*}
 \begin{center}
  \leavevmode
  \epsfxsize=15cm
  \epsffile{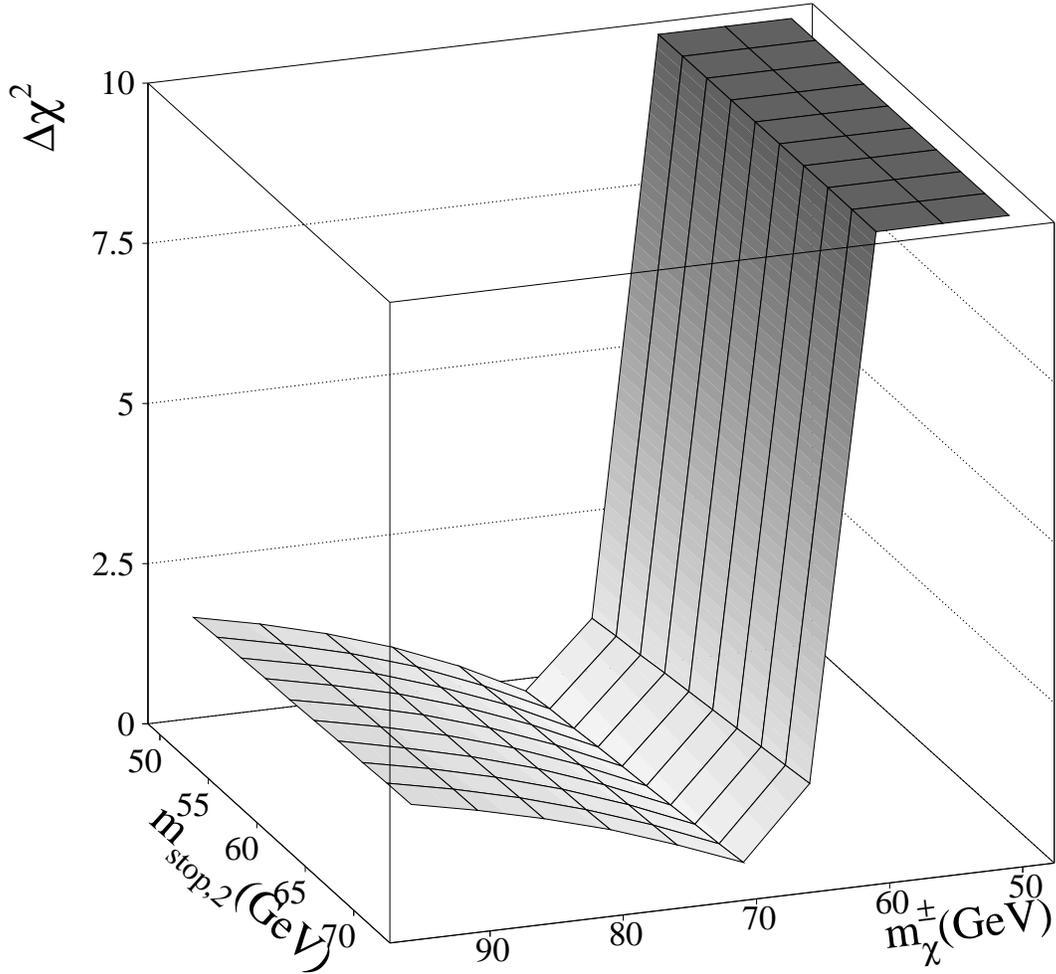}
 \end{center}
\caption{\label{\figochi}
The $\Delta\chi^2$ in the  light stop and light
chargino plane for $\tan\beta=1.6$. At each point of the grid an
optimization of $m_t$,$\alpha_s$
and the stop mixing angle
$\phi_{mix}$ was performed with $\mu > -40$ and $M_2=3\mid\mu\mid$, including
the ratio  $b\rightarrow s\gamma$. 
} 
\end{figure*}

\begin{figure*}
 \begin{center}
  \leavevmode
  \epsfxsize=15cm
  \epsffile{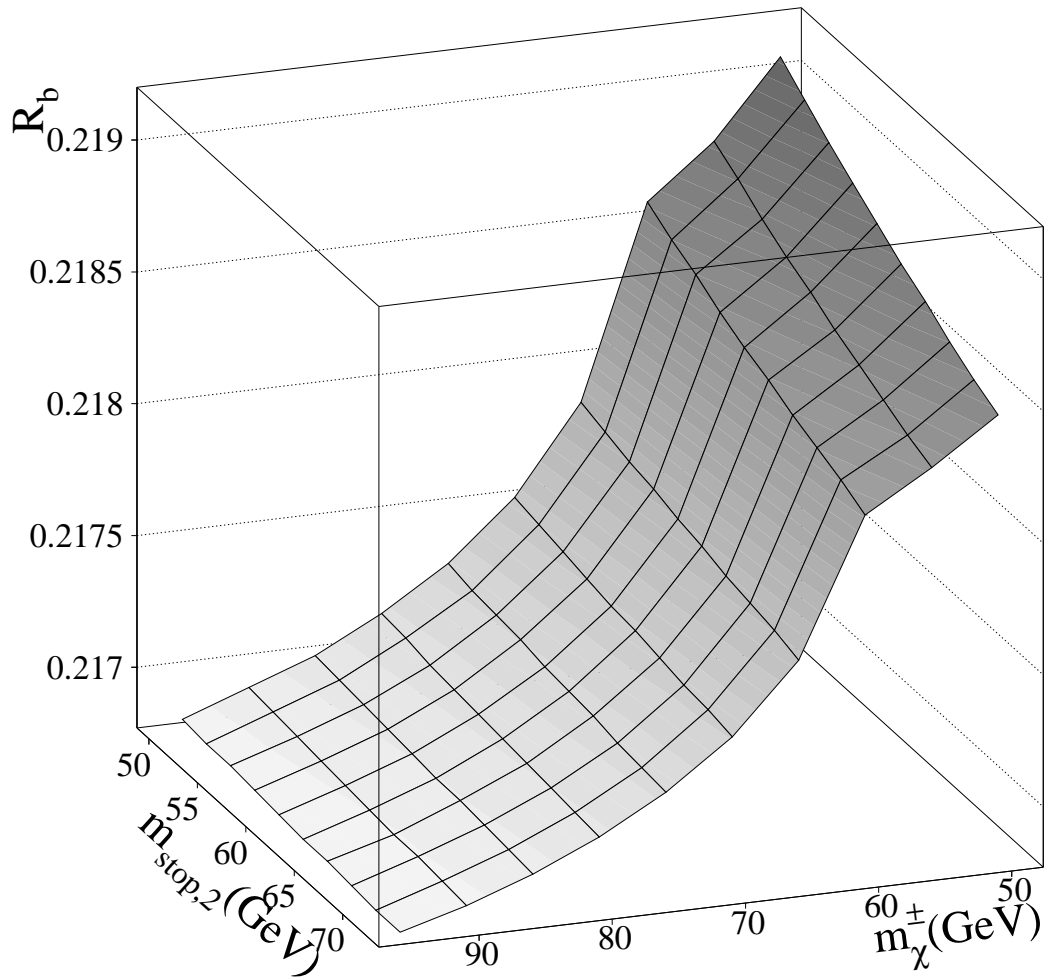}
 \end{center}
\caption{\label{\figorb}
$R_b$ in the  light stop and light
chargino plane. Optimization as in  fig.~\protect\ref{\figochi}.
} 
\end{figure*}

\begin{figure*}
 \begin{center}
  \leavevmode
  \epsfxsize=15cm
  \epsffile{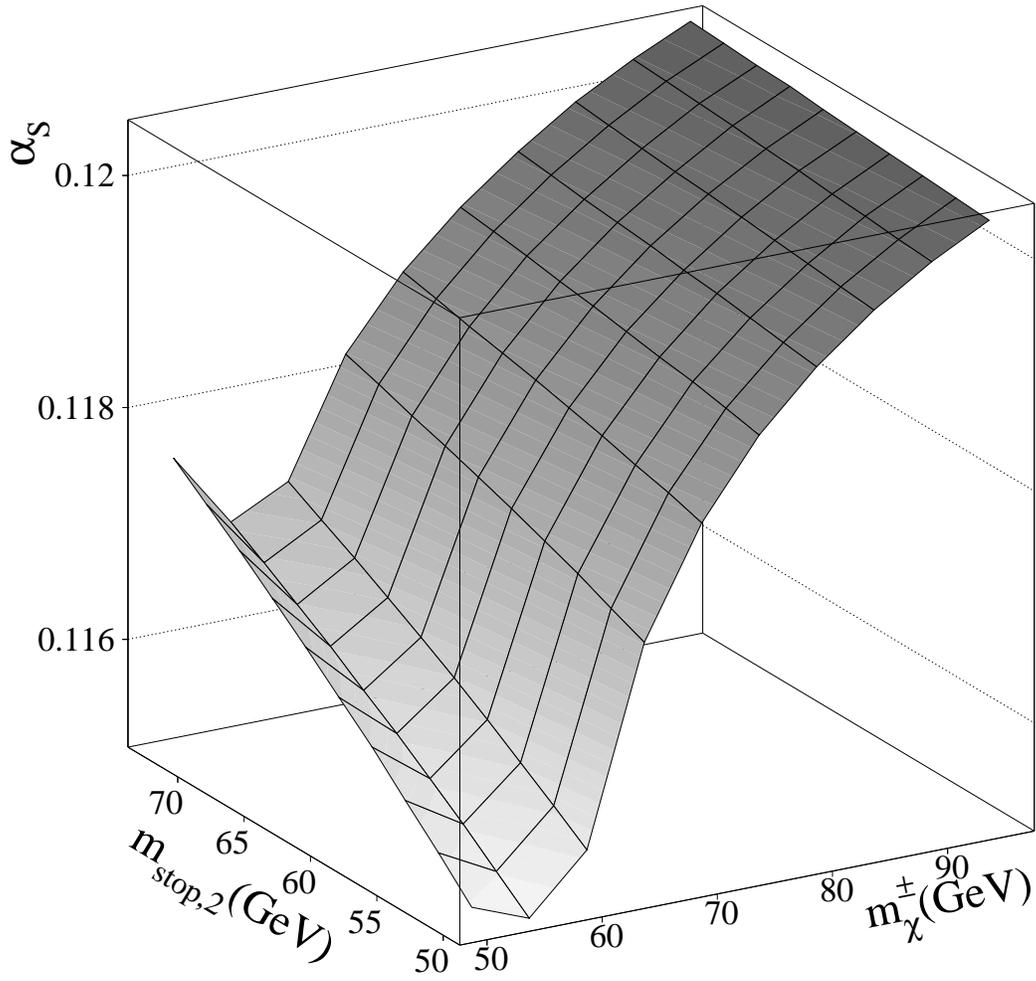}
 \end{center}
\caption{\label{\figoalf}
$\alpha_s$ in the  light stop and light
chargino plane. Optimization as in  fig.~\protect\ref{\figochi}.
} 
\end{figure*}

\begin{figure*}
 \begin{center}
  \leavevmode
  \epsfxsize=15cm
  \epsffile{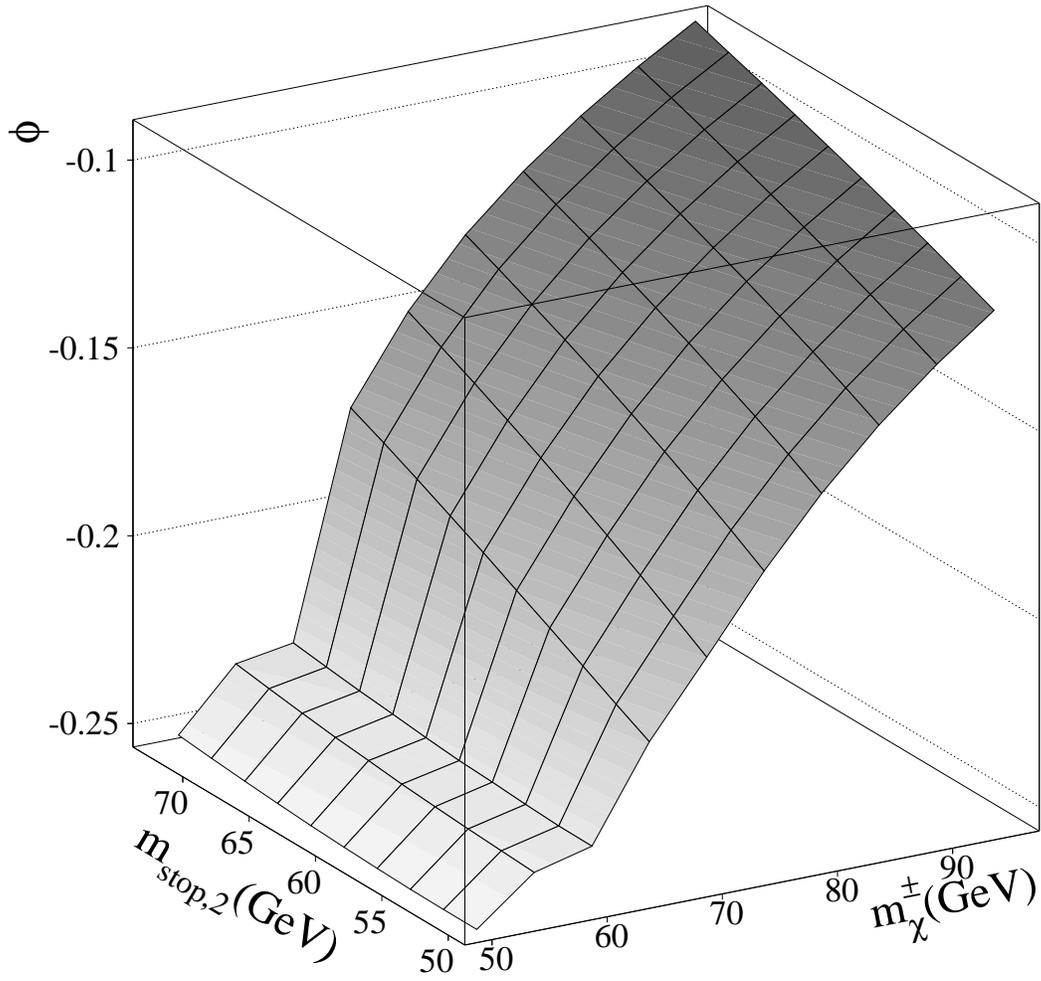}
 \end{center}
\caption{\label{\figomix}
Stop mixing angle $\phi_{mix}$ in the  light stop and light
chargino plane. It is mainly determined by the branching ratio  $b\rightarrow s\gamma$.
 Optimization as in  fig.~\protect\ref{\figochi}.
} 
\end{figure*}

\begin{figure*}
 \begin{center}
  \leavevmode
  \epsfxsize=15cm
  \epsffile{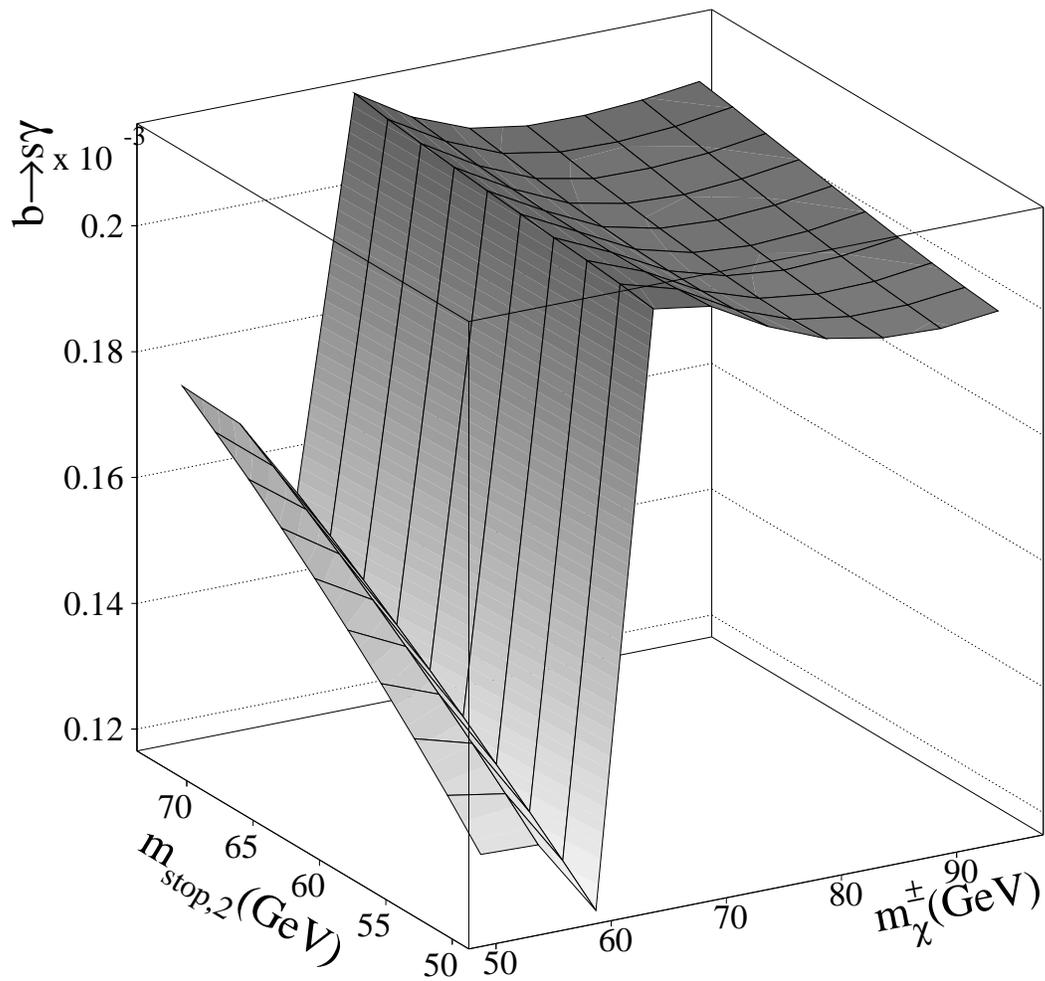}
 \end{center}
\caption{\label{\figobsg}
$b\rightarrow s\gamma$ in the  light stop and light
chargino plane. For chargino masses
higher than 60~GeV (and $\mu > 0 $) the predicted
value is close to 
the CLEO measurement of
$2.32\pm0.67\times 10^{-4}$.
Optimization as in  fig.~\protect\ref{\figochi}.
} 
\end{figure*}

\begin{itemize}
\item{Low $\tan\beta$ :}\\
\vspace{0.1cm}

Fig.~\ref{\figochi} shows the change in the best obtainable $\chi^2$ in the chargino - stop plane.
For each value of the lighter scalar top $m_{\tilde t_{2}}$ and $m_{\tilde \chi_2^\pm}$ in a grid of 10$\times$10 points
an optimization of $m_t$, $\alpha_s$ and the stop mixing angle $\Phi_{mix}$ was performed,
assuming $M_2=3\mid\mu\mid$ for a fixed value of $\tan\beta=$1.6. 
In the next section this assumption will be dropped.
Low sparticle masses yield a
sharp increase in the $\Delta\chi^2$ in fig.~\ref{\figochi} because of the included mass limits.
The minimum  $\chi^2$  is
obtained for chargino masses just above the experimental limit, although
it increases only slowly with increasing sparticle masses.
$R_b$ increases significantly with decreasing values of
the stop and chargino mass, as can be seen from fig.~\ref{\figorb}.
Much less significant is the improvement of
$R_c$. Within the plane of fig.~\ref{\figorb} it changes less than 0.0005 units.
The increase of $R_b$ must be compensated by a decrease  of $\alpha_s$ (see fig.~\ref{\figoalf})
in order to keep the total  Z-width constant.
The stop mixing angle $\Phi_{mix}$, shown in fig.~\ref{\figomix}, is mainly determined by the CLEO
measurement of
$b\rightarrow s\gamma$.
The chargino contribution to $b\rightarrow s\gamma$ is proportional to the Higgs mixing parameter
$\mu$, which changes its sign for $m_{\tilde\chi^{\pm}}\approx$ 60~GeV (see fig.\ref{\figI}), so the
$b\rightarrow s\gamma$ rate changes rapidly for these chargino masses, as shown in  fig.~\ref{\figobsg}.
The uncertainty in the
predicted $b\rightarrow s\gamma$ rate from the renormalization scale 
 has been taken into account and was varied between $m_b/2$ and
$2m_b$~\cite{buras}.
The scale itself has been chosen to be $m_b$.\\

\item{High $\tan\beta$:}\\
\vspace{0.1cm}

Similar fits can be performed for the high $\tan\beta$ scenario  in the pseudo scalar Higgs $m_A$ versus
light chargino plane.
As in the low $\tan\beta$ case $M_2=3\mid\mu\mid$ was
assumed. In fig.~\ref{\fighochi}  the resulting change in the $\chi^2$ is
given for fixed $\tan\beta=50$. For small chargino masses there is a sharp increase in the
$\chi^2$ due to the corresponding mass limit, see above. The best values for $R_b$ can be
obtained for small values of $m_A$ and $m_{\tilde\chi^{\pm}}$, see fig.~\ref{\fighorb}.
As in the low $\tan\beta$ case
the enhancement of $R_b$ must be compensated by a decrease of $\alpha_s$,
see fig.~\ref{\fighoalf}, and
the improvement of $R_c$ is small, less than 0.0006 within the given
parameter plane. 
The mixing angle,shown in fig.~\ref{\fighomix}, is mainly determined by the $b\rightarrow s\gamma$ rate,
which can be fitted in the whole  $m_A$ - chargino plane, see fig.~\ref{\fighobsg}.
\end{itemize}

\subsection{Best Solutions}

{\it Standard Model Fits:}\\
\vspace{0.1cm}

The input values from $M_W$ , $m_t$, the electroweak mixing angle and the $Z^0$ line shape observables
have been summarized in table \ref{fitresults}. The SM predictions were obtained from the ZFITTER
package \cite{zfitter} and all the error correlations were taken from \cite{Lep2}.
The fits were made with  $\alpha_s$, $m_t$ and
$m_H$ as free parameters, which resulted in
\begin{eqnarray*}
\alpha_s=0.1215\pm0.0036\\
m_t=167.3^{+8.2}_{-7.6}~{\rm GeV}\\
m_H=66^{+81}_{-37}~{\rm GeV}.
\end{eqnarray*}
The quoted errors have been determined using MINOS~\cite{minuit}.
Further details of the  procedure are described  elsewhere, see for example~\cite{ralf,cmssm,wimhig}.
The $\chi^2/d.o.f$ of
the SM fit is 23.2/15 which corresponds to a probability of 8\%. Here, the main contributions to the
$\chi^2$ originate from $R_b$ ($\Delta\chi^2=10.7$), $\sin^2\Theta_{eff}^{lept}$ from SLD  ($\Delta\chi^2=3.6$)
and $R_c$  ($\Delta\chi^2=3.3$).
The correlation parameter
between $m_H$ and $m_t$ for the best fit is approximately 0.7; this strong correlation is shown in fig. \ref{\mtmh}.
One observes that the upper limit on the Higgs mass is obtained for $m_t\approx 175$~GeV; however, the
upper limit is sensitive to $\sin^2\Theta_{eff}^{lept}$ as shown by the dashed contour in fig.~\ref{\mtmh}, where the
precise value of $\sin^2\Theta_{eff}^{lept}$ from SLD was excluded from the fit.
The  dependence of $\sin^2\Theta_{eff}^{lept}$ on the SM Higgs mass is approximately logarithmic
(see fig. \ref{\sintw}).
The LEP data alone without SLD yields $m_H=144^{+164}_{-82}$~GeV, SLD alone yields $m_H=15^{+25}_{-8}$~GeV,
as indicated by the squares in fig. \ref{\sintw}.
The latter value is excluded by the lower limit of 58.4~GeV from the combined LEP experiments~\cite{rev96}.
In addition, the ALEPH Collaboration gives a recent limit of 63.9~GeV on the SM Higgs mass \cite{lim6}. 
The $\Delta\chi^2$ dependence of the Higgs mass is shown in fig. \ref{\smdchi} for various conditions.\\

\vspace{0.5cm}

{\it MSSM Fits and Comparison with the SM:}\\

In order to obtain the best MSSM fits the assumption $M_2=3\mid\mu\mid$ is dropped
and $M_2$ is treated as a free parameter.
$R_b$ increases with decreasing $\tan\beta$. The fit results for a $\tan\beta$ as low as one
are given in the first column of table \ref{bestfit}, the corresponding predictions of electroweak
observables in table \ref{fitresults}.
Note the high value of $R_b$ and the corresponding
low value of $\alpha_s$.
The resulting $\chi^2/d.o.f.$ is 15.1/11,
corresponding to a probability of about 18\%.
Unfortunately, the corresponding $b\rightarrow s\gamma$ rate is about one order of magnitude too
small in this region of parameter space.
Larger rates can be obtained either by heavier sparticle masses or by larger values of $\tan\beta$.
With  $b\rightarrow s\gamma$ included in the fit and a free $\tan\beta$, the preferred
value of $\tan\beta$ was either around 1.6 or 50.

%The parameter optimization has been performed for $\tan\beta=1.6$ and $\tan\beta=50$ with all
%limits and constraints mentioned above included. 
The fit results for these values of $\tan\beta$ are given in table
\ref{bestfit} too and the predicted values of all observables with their pulls have
been summarized in table \ref{fitresults}. Note that
the MSSM prediction of the W-boson mass is always higher
than the Standard Model one.

For the best solutions the gluino mass 
was fixed to 1500~GeV, the stau mass
to 500~GeV and the  sbottom mass to 1000~GeV in both the low and the high $\tan\beta$
scenario, since they 
are less sensitive to the LEP observables and therefore cannot be fitted.
Their influence was studied by
fixing them to different values and repeating the fits again. First  the low $\tan\beta$ scenario
will be discussed.
A variation of the gluino mass
%within the low $\tan\beta$ scenario 
from 200~GeV up to 2000~GeV did not change the best obtainable $\chi^2$, varying the stau
mass in the same range changed the $\chi^2$ less than 0.2. Furthermore, the best obtainable
$\chi^2$ changed less than 0.2 when varying the
sbottom mass and pseudoscalar Higgs mass from 800~GeV to 2000~GeV. For values of these two parameters
below 800~GeV the $\chi^2$ increased significantly, mainly because 
the prediction of $R_b$ became too small.
%However, if they were
%chosen heavy enough, the dependence on them vanished.

Within the high $\tan\beta$ scenario no significant
change of the global $\chi^2$ was detected when the gluino mass was varied between 200~GeV and 2000~GeV,
but the stau mass was somewhat more sensitive. A variation of this parameter between
 300~GeV and 700~GeV  changed the global $\chi^2$ less than 0.2, but if it was chosen
higher than 1000~GeV the $\chi^2$ increased up to 1.6 units, mainly because
the prediction of ${\cal A}_\tau$ became worse.
The sbottom mass was varied between 800~GeV and 2000~GeV. As in the low $\tan\beta$ case
there was no dependence on this parameter if it was chosen heavy, but for low values the
preferred top mass became too small.
$M_2$ was fixed at 1500~GeV. A variation
between 1000~GeV and 2000~GeV changed the best reachable $\chi^2$ less than 0.2, for
smaller values the global $\chi^2$ increased up to one unit.

To check the influence of the assumptions on $M_1$ for the best fits, 
the GUT relation
$M_1 = 5/3 \tan^2 \theta_W \, M_2$
was dropped and the fit was repeated with a free $M_1$, but this
did not improve the best obtainable $\chi^2$ significantly.

A direct comparison to the Standard Model fits is given for all three fits in
figs.~\ref{\figIV}-\ref{\figVI}. The resulting Standard Model
$\chi^2/d.o.f.=23.2/15$ corresponds to a probability of 8\%,
the MSSM fits
correspond to probabilities of 15\% ($\tan\beta=1.6$, $\chi^2/d.o.f.=16.9/12$) and 10\%
($\tan\beta$=50, $\chi^2/d.o.f.=18.4/12$).
In counting the d.o.f the insensitive (and fixed) parameters were ignored.
The high $\tan\beta$ scenario is hardly more probable than the
SM, while for low $\tan\beta$ the probability of 15\% is in between the best MSSM fit
without $b\rightarrow s\gamma$  and the SM fit.

Another interesting point are the predictions for  $\alpha_s(M_Z)$.
Fig.~\ref{\alphas} shows a comparison of different measurements
of $\alpha_s(M_Z)$ (as given in \cite{rev96,lattbb}) with the fitted values given in this paper.
The fitted MSSM $\alpha_s(M_Z)$ is slightly smaller than the Standard Model value and
 in better agreement with measurements from deep inelastic lepton scattering (DIS)
and the world average\cite{rev96}.

%Quite important constraints are the limits on the lightest and second lightest neutralino.

\subsection{Discovery Potential at LEP II}
The particle spectrum for the best fits, as shown in table \ref{bestfit},
suggests that some SUSY particles could be within  reach of LEP II.
If they are not found, LEP II will provide stringent SUSY mass limits\cite{cmssm,wimhig}.
In the
following the consequences  of increased mass limits on the fits are discussed for
both the low and high $\tan\beta$ scenario.\\ 

{\it Chargino Searches:}\\
\vspace{0.1cm}

\begin{figure*}
 \begin{center}
  \leavevmode
  \epsfxsize=15cm
  \epsffile{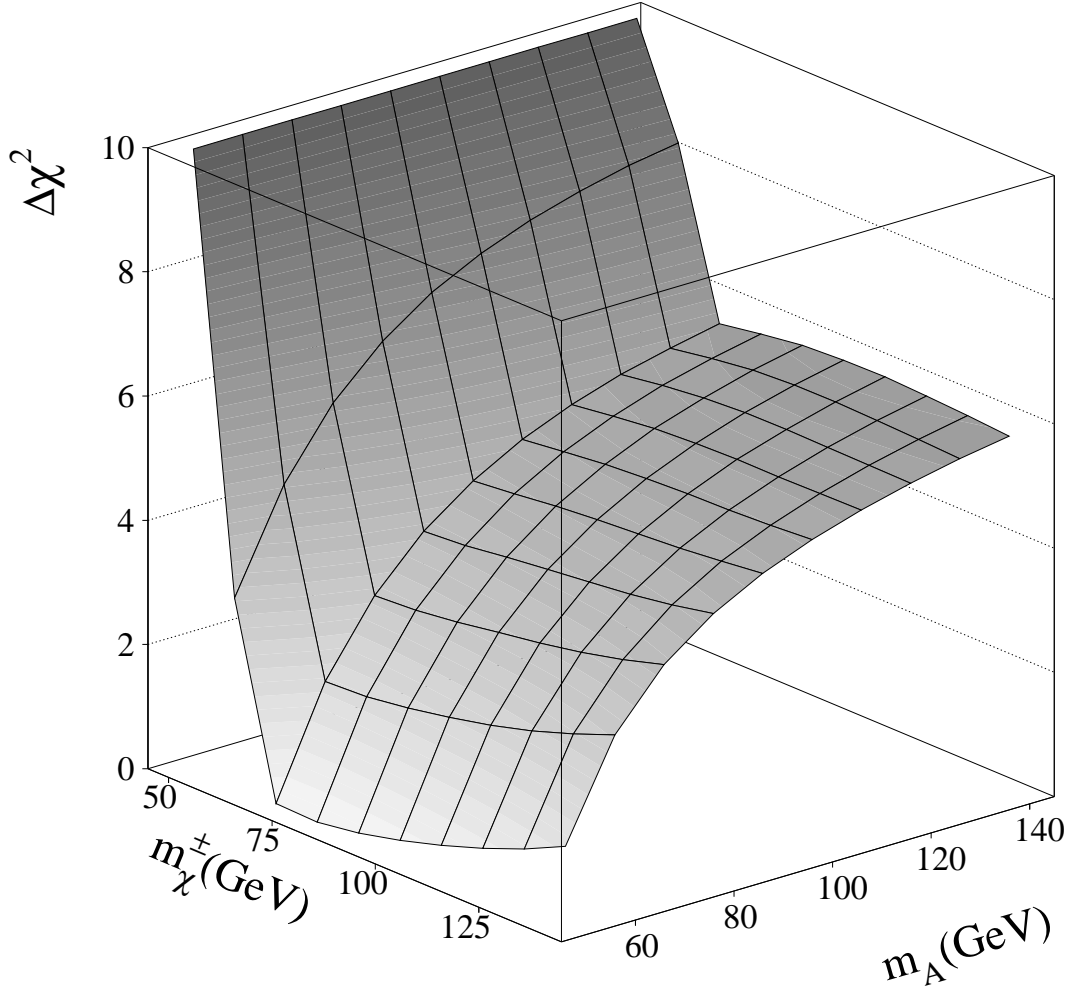}
 \end{center}
\caption{\label{\fighochi}
The $\Delta\chi^2$ in the pseudo scalar Higgs and light chargino plane for $\tan\beta=50$.
For each given $m_A$ and light 
chargino mass an optimization of $m_t$ , $\alpha_s$ , $\tilde{m}_{t2}$ and the stop
mixing angle $\phi_{mix}$ was performed, including the $b\rightarrow s\gamma$ rate and with
the irrelevant parameter $M_2$ set to 1500~GeV. 
} 
\end{figure*}

\begin{figure*}
 \begin{center}
  \leavevmode
  \epsfxsize=15cm
  \epsffile{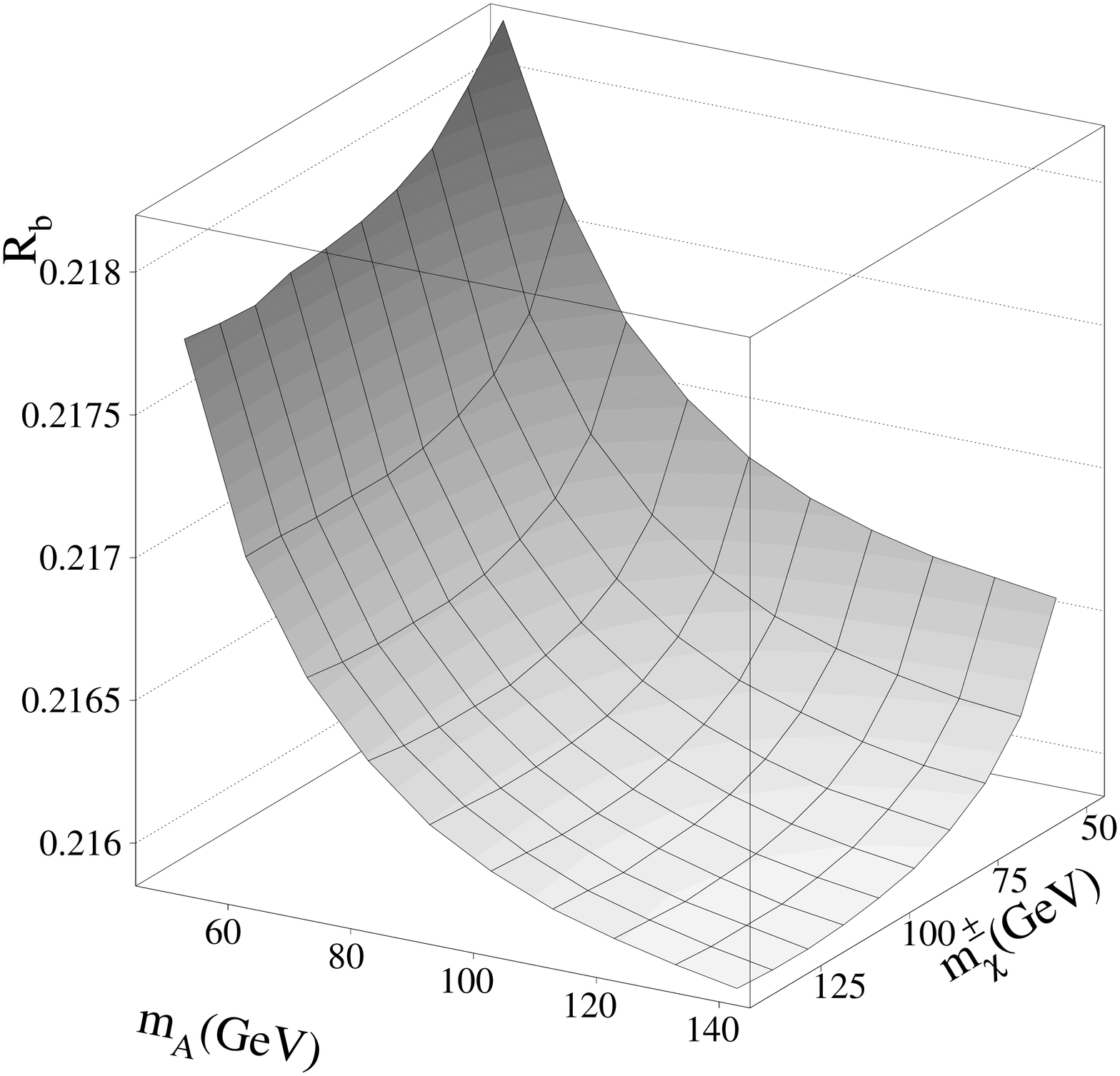}
 \end{center}
\caption{\label{\fighorb}
$R_b$ in the pseudo scalar Higgs and light chargino plane for $\tan\beta=50$.
 Optimization as in  fig.~\protect\ref{\fighochi}.
 } 
\end{figure*}

\begin{figure*}
 \begin{center}
  \leavevmode
  \epsfxsize=15cm
  \epsffile{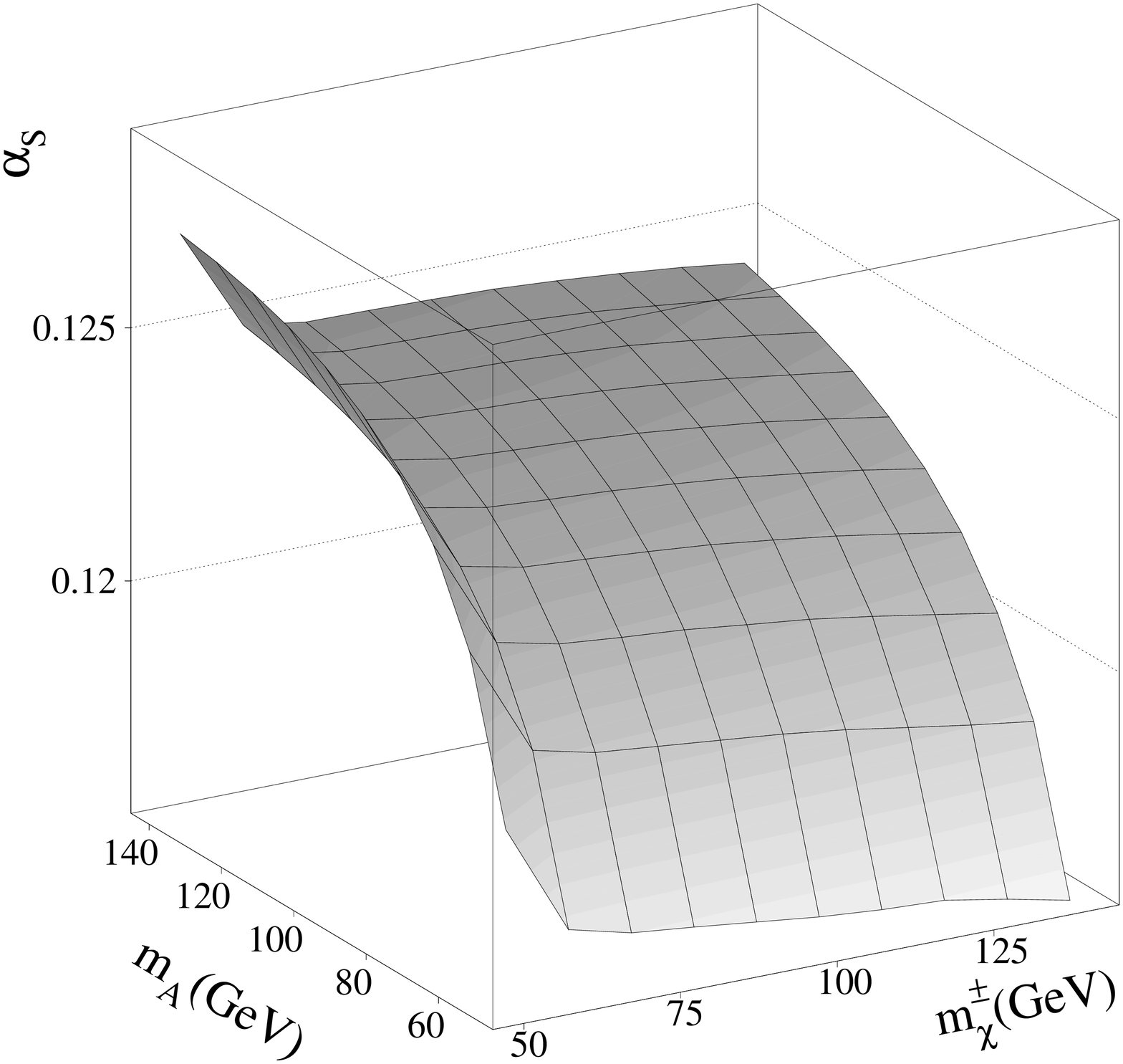}
 \end{center}
\caption{\label{\fighoalf}
$\alpha_s$ in the pseudo scalar Higgs and light chargino plane for $\tan\beta=50$.
Optimization as in  fig.~\protect\ref{\fighochi}.
} 
\end{figure*}

\begin{figure*}
 \begin{center}
  \leavevmode
  \epsfxsize=15cm
  \epsffile{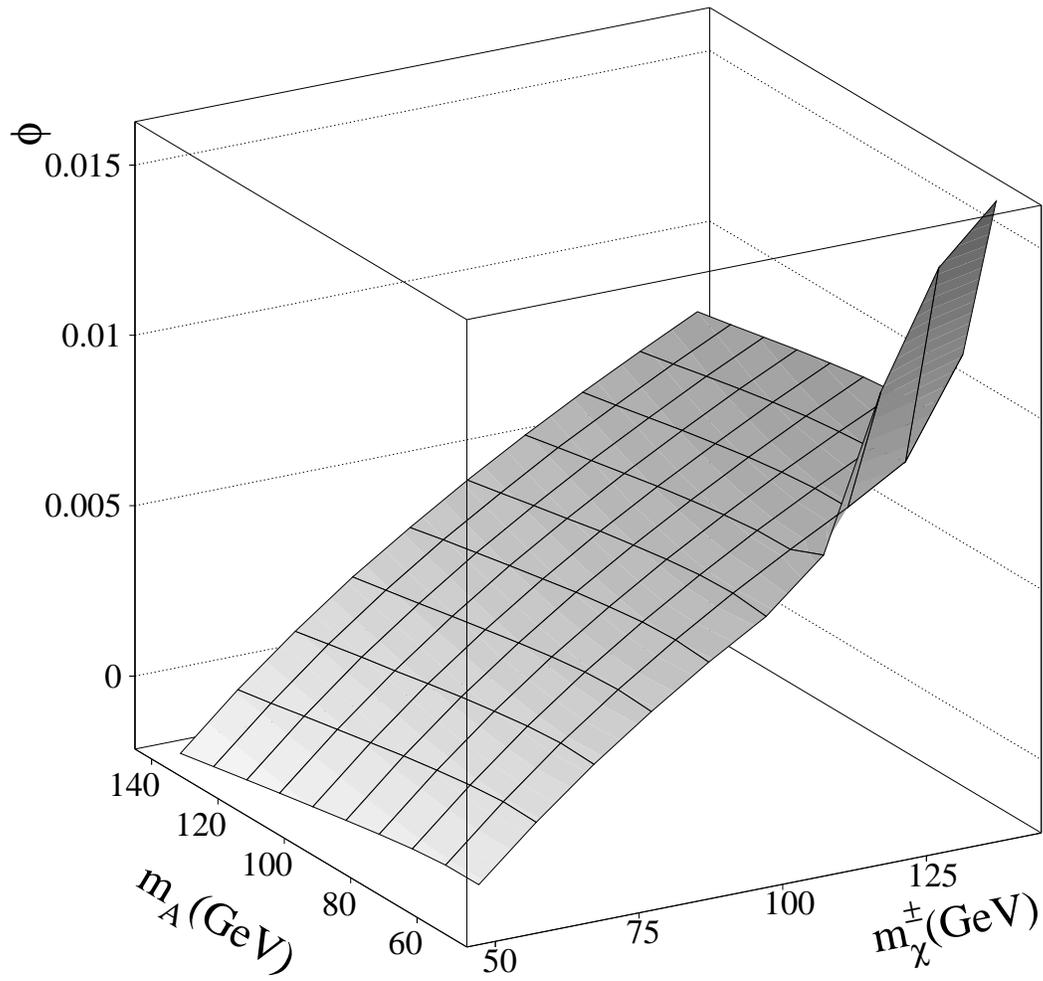}
 \end{center}
\caption{\label{\fighomix}
Stop mixing angle $\phi_{mix}$  in the pseudo scalar Higgs
and light chargino plane for $\tan\beta=50$.
Optimization as in  fig.~\protect\ref{\fighochi}.
} 
\end{figure*}

\begin{figure*}
 \begin{center}
  \leavevmode
  \epsfxsize=15cm
  \epsffile{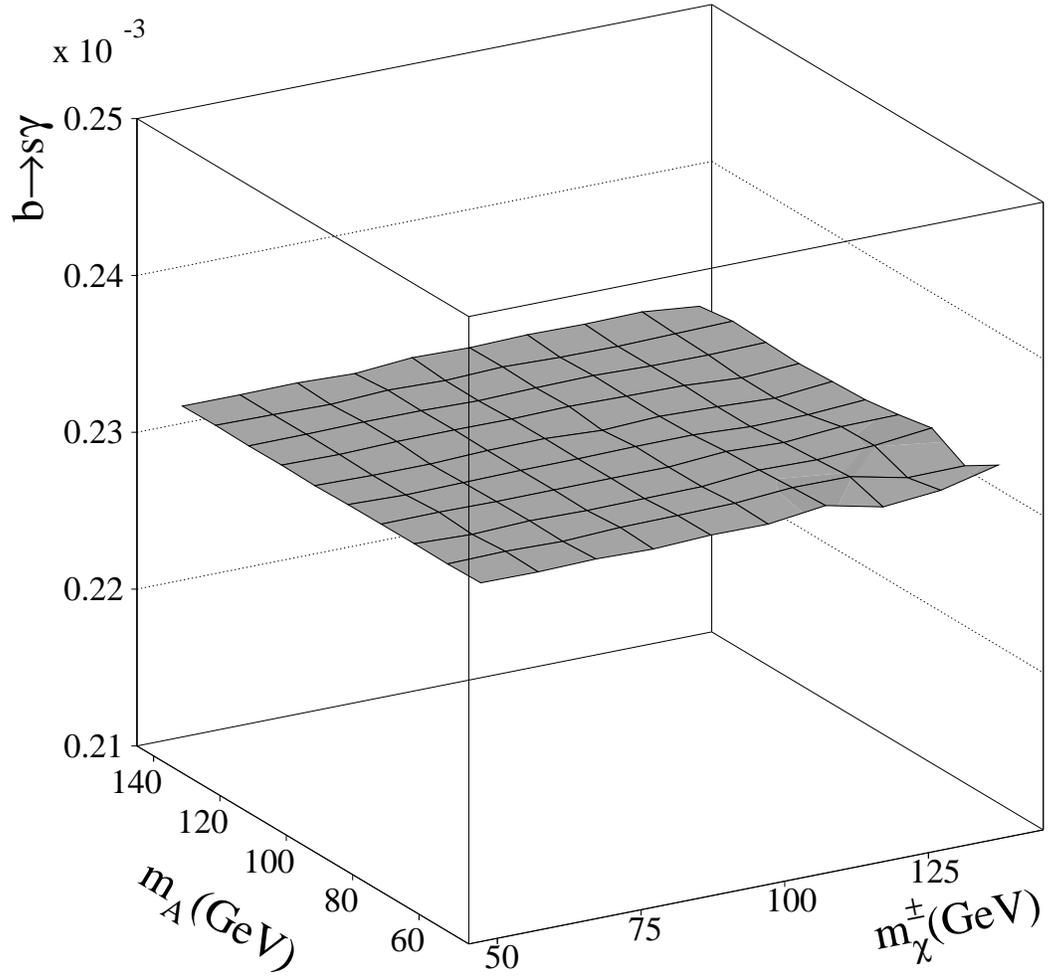}
 \end{center}
\caption{\label{\fighobsg}
$b\rightarrow s\gamma$ in the pseudo scalar Higgs
and light chargino plane for $\tan\beta=50$.
The prediction is close to the
CLEO measurement of 
$2.32\pm0.67\times 10^{-4}$
within the whole parameter space.
Optimization as in  fig.~\protect\ref{\fighochi}.
} 
\end{figure*}

\begin{figure*}
 \begin{center}
  \leavevmode
  \epsfxsize=15cm
  \epsffile{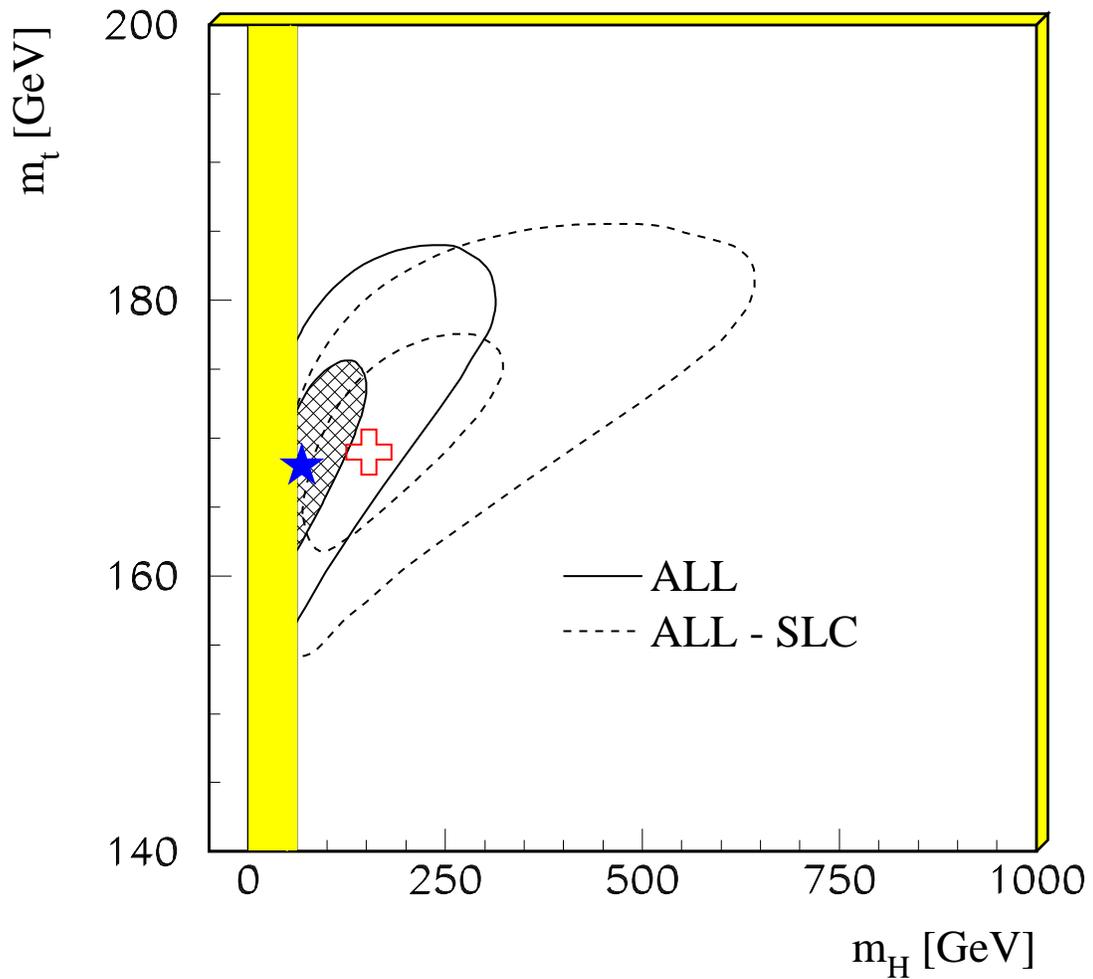}
 \end{center}
\caption{\label{\mtmh}
$\Delta\chi^2=1$ and $\Delta\chi^2=4$ contour lines for all electroweak data
including  $\sin^2\Theta_{eff}^{lept}$ from SLD (continous line) and without it
(dashed line). The stars indicate the best fits.
} 
\end{figure*}

\begin{figure*}
 \begin{center}
  \leavevmode
  \epsfxsize=15cm
  \epsffile{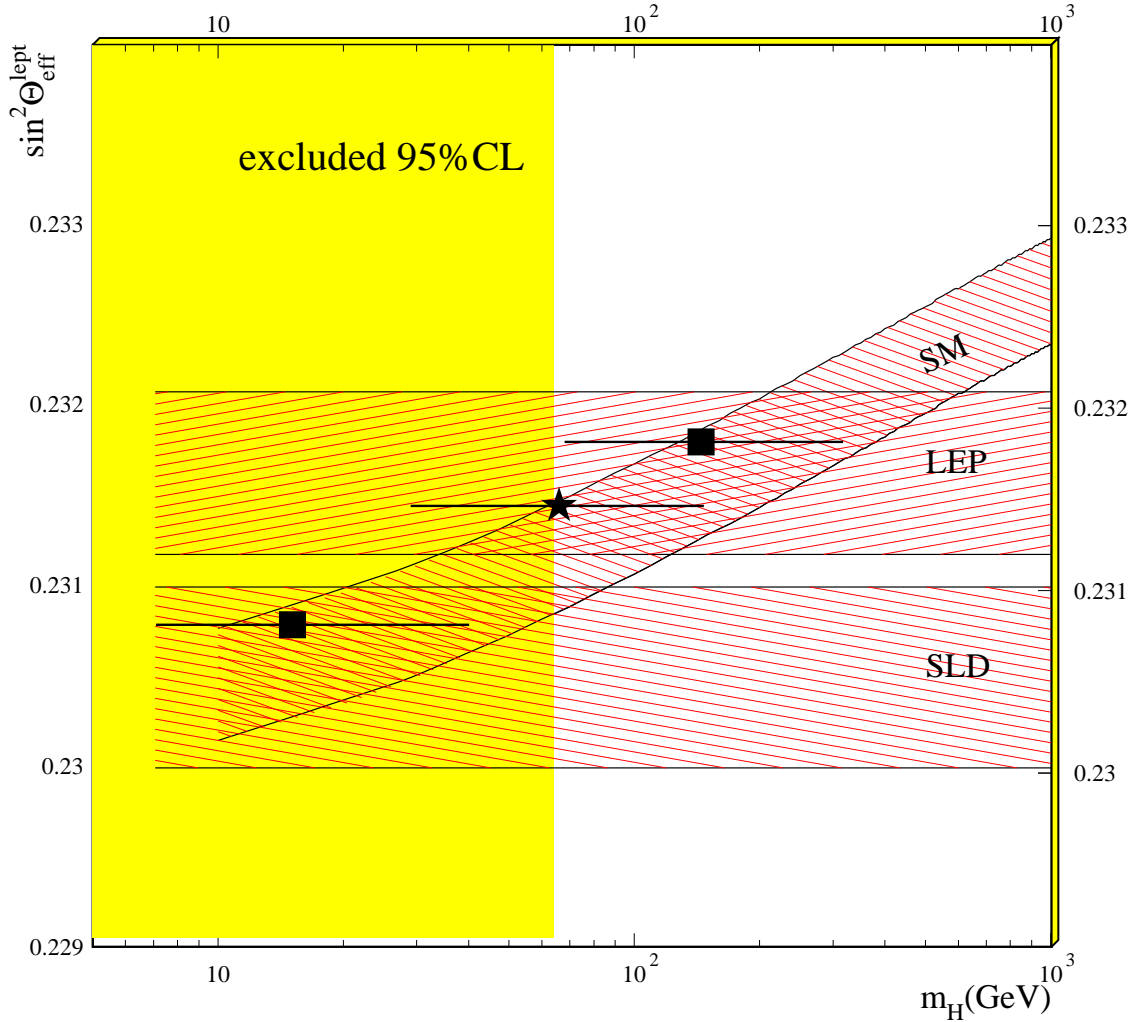}
 \end{center}
\caption{\label{\sintw}
Dependence of the SM $\sin^2\Theta_{eff}^{lept}$ on the Higgs mass. The top
mass $m_t=175\pm 9$~GeV was varied within its error, as shown by the dashed band
labelled SM (upper (lower) boundary $m_t$=166(184)~GeV). 
The SLD and the LEP measurement of  $\sin^2\Theta_{eff}^{lept}$ 
are also shown as horizontal bands. Fits to the electroweak data prefer $m_t\approx170$~GeV
and light Higgs masses, as indicated by the squares for the separate LEP and SLD measurements, while
the star is the result of the combined fit to SLD and LEP data. Clearly,
the SLD value yields a Higgs mass less than the recents limits of 63.9~GeV by direct Higgs searches at LEP
(shaded area)\protect\cite{lim6}.
} 
\end{figure*}

\begin{figure*}
 \begin{center}
  \leavevmode
  \epsfxsize=15cm
  \epsffile{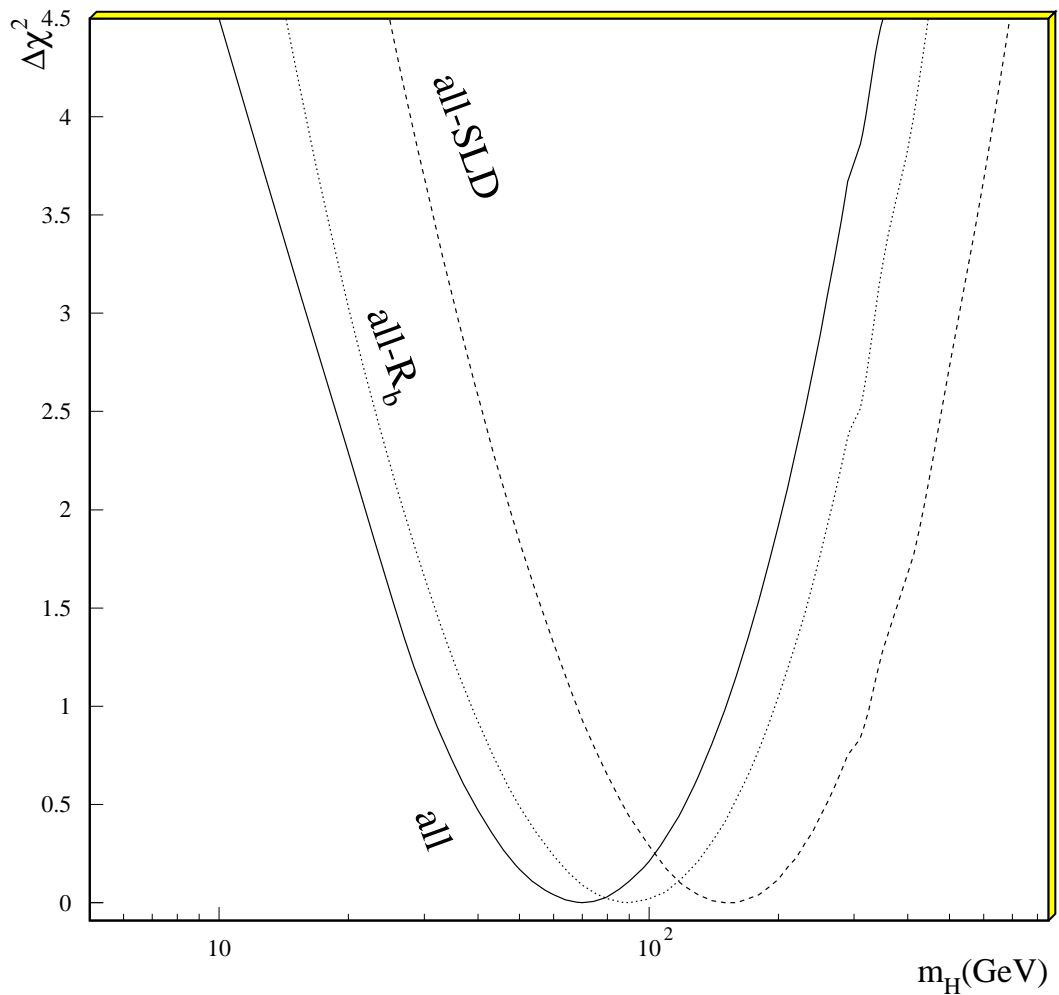}
 \end{center}
\caption{\label{\smdchi}
Dependence of the SM $\Delta\chi^2$ on the Higgs mass for a free top mass,
taking all data (continous line), all data
without the SLD measurement of  $\sin^2\Theta_{eff}^{lept}$ (dashed line) and
all data without $R_b$ (dotted line). 
} 
\end{figure*}

\begin{figure*}
 \begin{center}
  \leavevmode
  \epsfxsize=15cm
  \epsffile{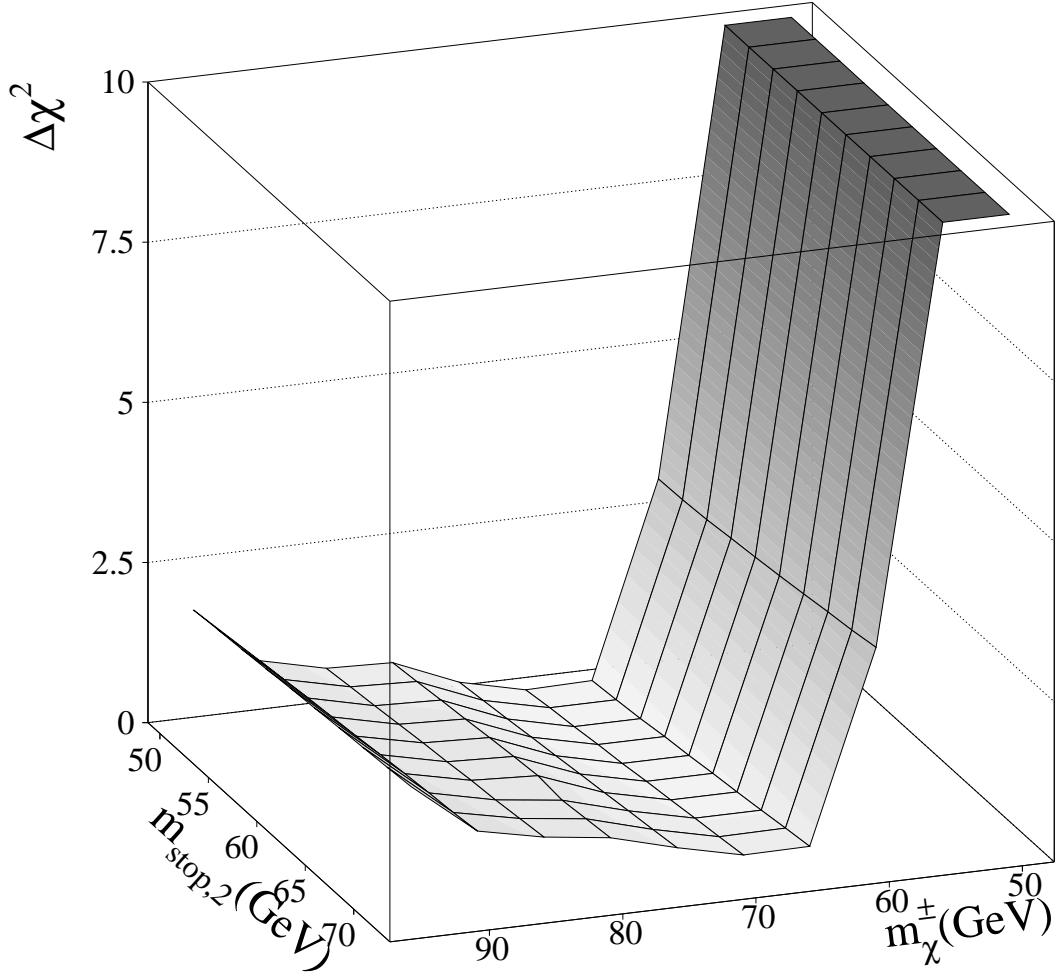}
 \end{center}
\caption{\label{\figochiall}
The $\Delta\chi^2$ in the region of the best fit in the  light stop and light
chargino plane for $\tan\beta=$1.6. Here the constraint on $M_2$ was dropped.
At each point of the grid an
optimization of $m_t$, $M_2$, $\alpha_s$
and the stop mixing angle $\phi_{mix}$ was performed with $\mu > -40$, including
the ratio  $b\rightarrow s\gamma$ and
the requirement $\Gamma_{Z\rightarrow neutralinos }< 2~$MeV. 
} 
\end{figure*}

\begin{figure*}
 \begin{center}
  \leavevmode
  \epsfxsize=15cm
  \epsffile{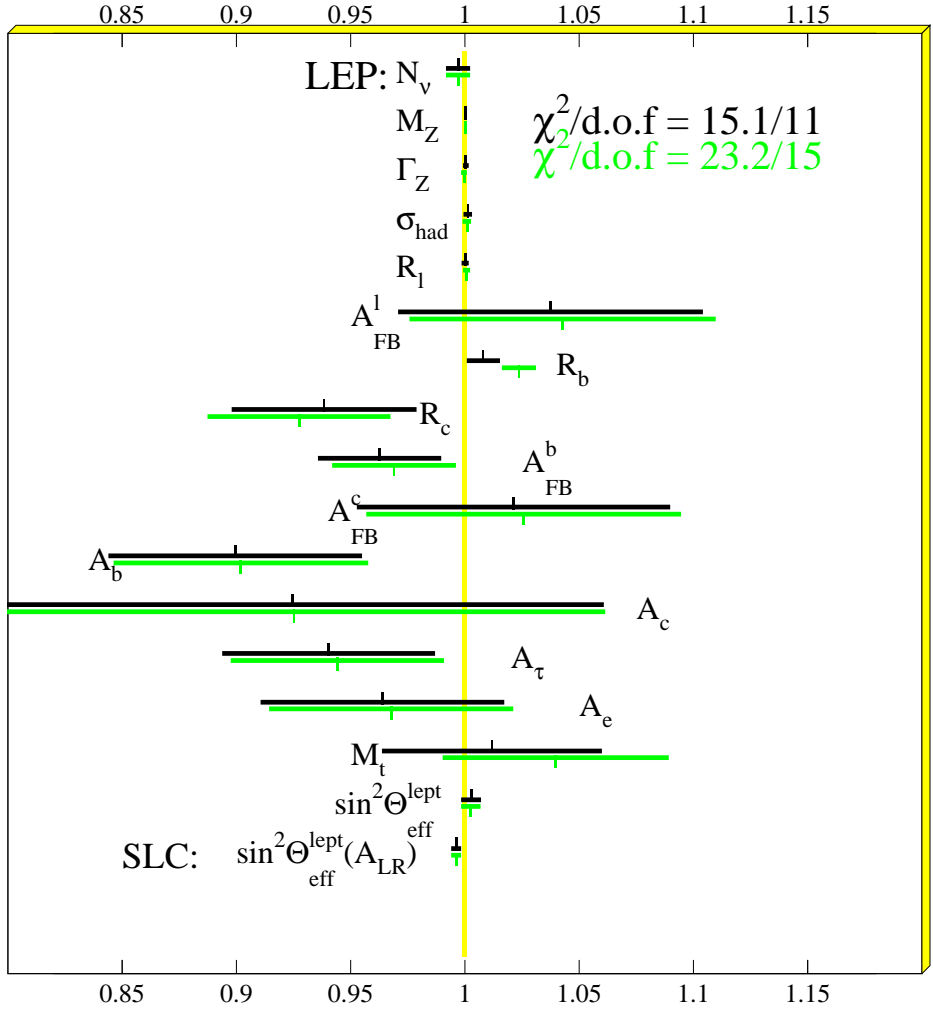}
 \end{center}
\caption{\label{\figIV}
Resulting observables for the fit given in table \protect\ref{bestfit} 
for $\tan\beta=1.0$.
A significant improvement of $R_b$ can be observed here. The ratio  $b\rightarrow s\gamma$
was not included in this fit here.} 
\end{figure*}

\begin{figure*}
 \begin{center}
  \leavevmode
  \epsfxsize=15cm
  \epsffile{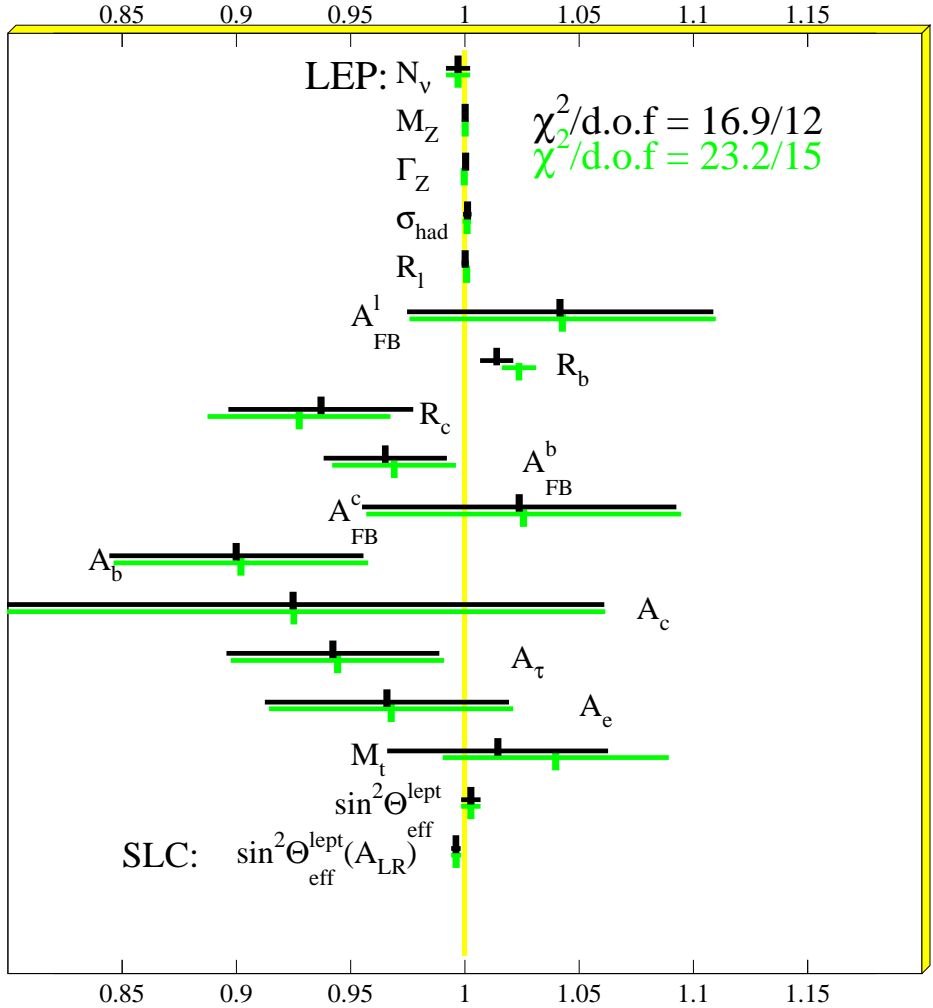}
 \end{center}
\caption{\label{\figV}
Resulting observables for the fit given in table \protect\ref{bestfit}
for $\tan\beta=1.6$. $\tilde{m}_b$ was
fixed to 1000~GeV, $m_A$ and the gluino mass were fixed to 1500~GeV.
Including $b\rightarrow s\gamma$ in the fit
it is still possible
to improve the prediction of $R_b$ with Supersymmetry a bit.}
\end{figure*}

\begin{figure*}
 \begin{center}
  \leavevmode
  \epsfxsize=15cm
  \epsffile{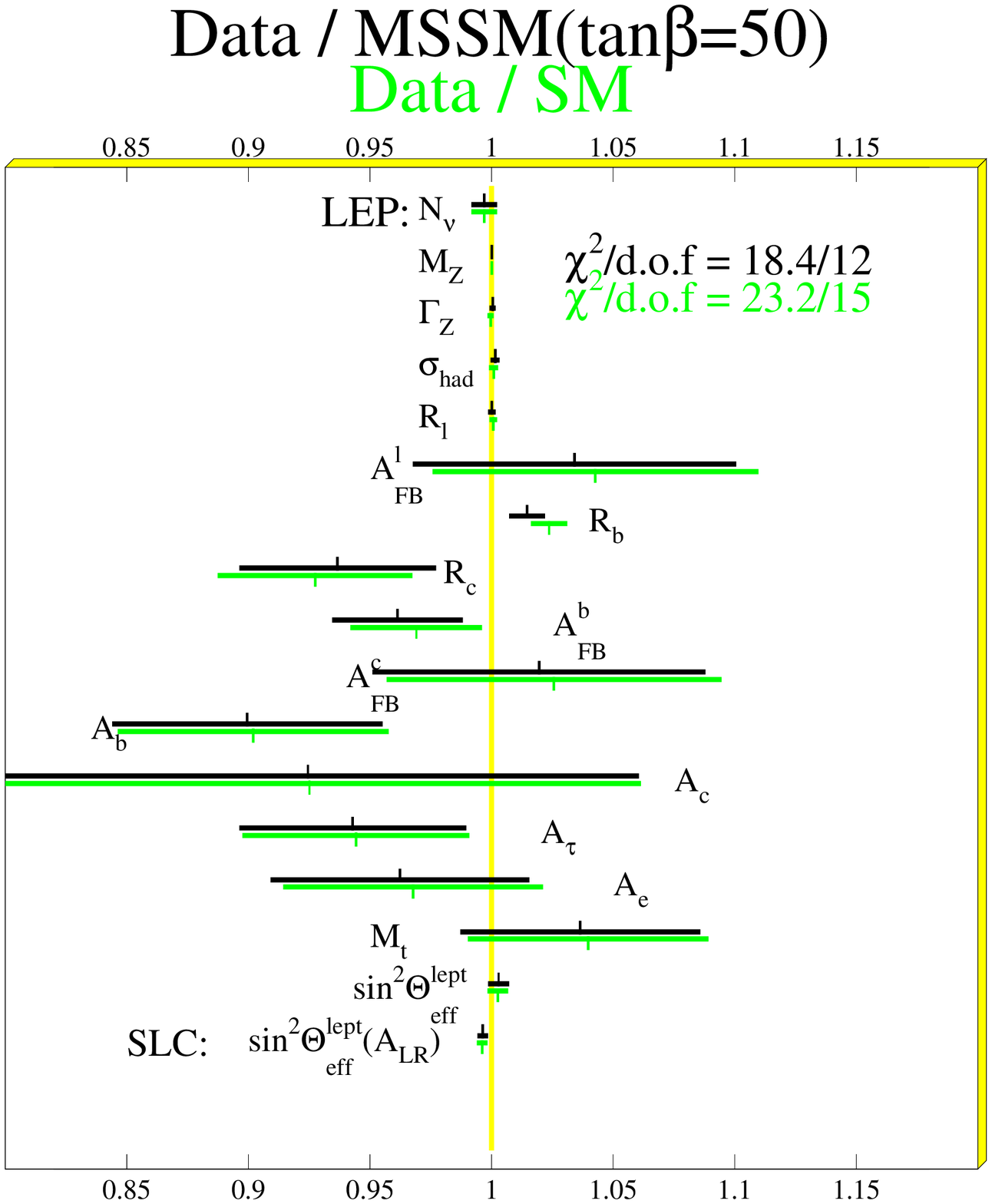}
 \end{center}
\caption{\label{\figVI}
Resulting observables for the fit given in table \protect\ref{bestfit}
for $\tan\beta=50$. $\tilde{m}_b$ was
fixed to 1000~GeV, $M_2$ and the gluino mass were fixed to 1500~GeV. It is possible
to improve the prediction of $R_b$ with Supersymmetry even for high values of
$\tan\beta$, but the result is not as good as for low values.}
\end{figure*}

\begin{figure*}
 \begin{center}
  \leavevmode
  \epsfxsize=15cm
  \epsffile{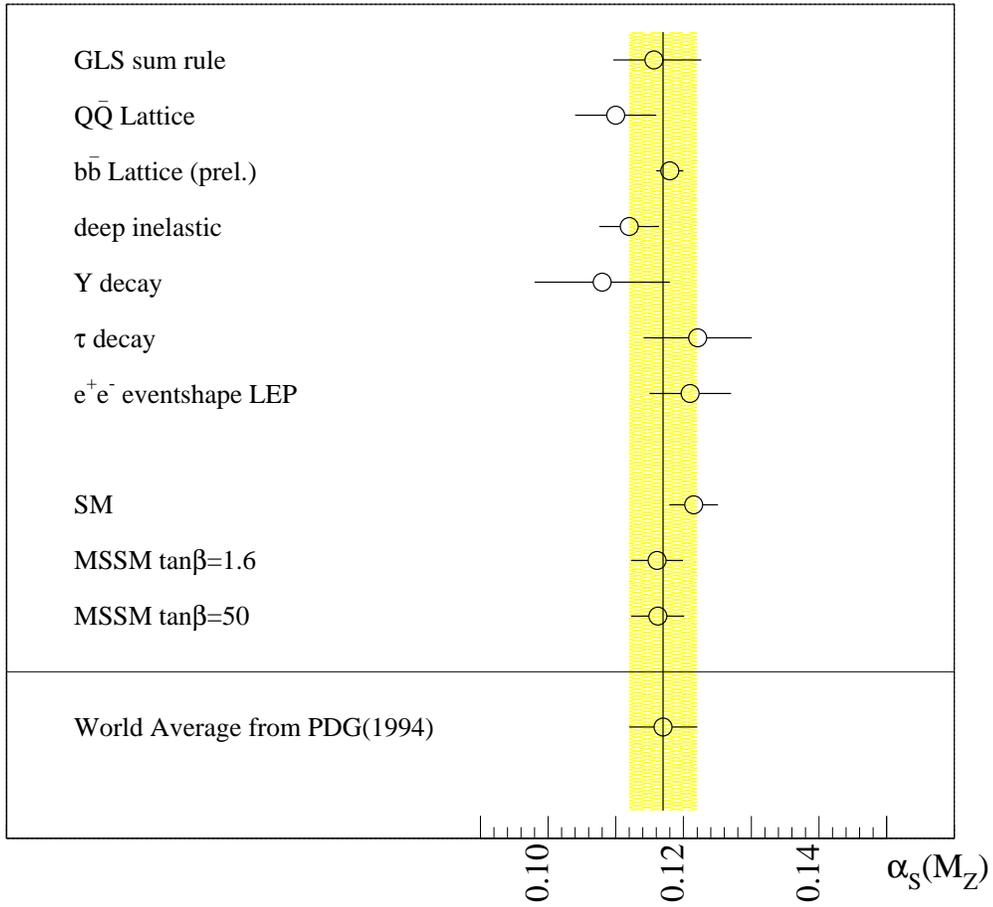}
 \end{center}
\caption{\label{\alphas}
Comparison of different measurements of $\alpha_s$ with the fit results.
The data has been taken from \protect\cite{rev96} and \protect\cite{lattbb}.}
\end{figure*}

The $\chi^2$
in the region of the best low $\tan\beta$ fit increases
slowly for increasing chargino masses, see fig.~\ref{\figochiall}.
Chargino masses above a possible LEP II
limit of 95~GeV will increase the global $\chi^2$ of the fit by approximately 2 units, which corresponds 
to a probability of 9\%, which is hardly better than the SM probability, so one cannot consider
the MSSM as a better solution in that case. For the high $\tan\beta$
scenario the dependence of the global $\chi^2$ on the chargino mass
is small, see fig.\ref{\fighochi}, so increased limits will hardly  change  the 
probability of about  10\% for  the best possible fit in this scenario.\\

{\it Stop and Neutralino Searches:}\\
\vspace{0.1cm}

The best $\tan\beta=1.6$ fit has a light stop mass of about 48~GeV, and the lightest
neutralino is about 20~GeV, so this solution is just in between the regions of the
parameter space which are excluded by stop searches at LEP1 and the Tevatron
~\cite{lim4}: LEP I limits for light right handed stops are about 45~GeV, while
D0 limits exclude $52 (70)~$GeV$ < m_{\tilde t_{2}} < 92 (87)$~GeV for
$m_{\tilde{\chi_{0}}}=$ 20 (40)~GeV.
A fit with  increased neutralino mass limits of 45~GeV  
yielded a best solution of $\chi^2/d.o.f. = 17.8/12$, corresponding to  a probability of
12\%, which is  worse than the solution presented above, but still better
than the Standard Model one. A similar  $\chi^2/d.o.f.$ was obtained if the light stop
was required to be heavier than 90~GeV. The reason for this can be found in the flat
$\chi^2$ distribution in fig.~\ref{\figochiall}.
This figure is similar to fig.~\ref{\figochi}, but here $M_2$ was treated as
a free input parameter.

The stop mass in the high $\tan\beta$ solution is 53~GeV, and the
lightest neutralino is quite heavy, above 70~GeV, so there is no conflict with
stop searches. Furthermore, the  high $\tan\beta$ is insensitive to the stop mass,
so an inreased mass limit does not effect the fit.\\
%Requiring a stop mass higher than 90~GeV within this scenario yielded a best
%$\chi^2/d.o.f$ of approximately 18.6/12 corresponding to a probability of 10\% which is
%comparable to the high  $\tan\beta$ solution given in table \ref{bestfit}. \\

{\it Higgs Searches:}\\
\vspace{0.1cm}

The mass of the light scalar Higgs
is a sensitive function of the top mass. If the top mass is below 180~GeV, the Higgs mass
for the  low $\tan\beta$ fit (see tab.~\ref{bestfit}) should be
observable at LEP II, especially if one includes the second order corrections, which will lower the
Higgs mass by 10 - 15 GeV~\cite{wimhig}.
Within the high $\tan\beta$ scenario both  neutral
  Higgs bosons are light and have practically the same mass. Increasing the Higgs limits  
 up to 90~GeV increases the $\chi^2/dof$ to 23.7/12, corresponding to a probability
of only 2\%, which is much worse than the SM fit.
The steep dependence of the  $\chi^2$  on the pseudoscalar Higgs mass, which is mainly
caused by a too small value of $R_b$, is shown in  fig.~\ref{\fighochi}.

\section{CMSSM and $R_b$}

In ref.~\cite{cmssm} fits to low energy data have been performed within the
constrained MSSM (CMSSM). In this case unification of gauge and
b-$\tau$ Yukawa couplings is assumed.
Reproducing the large mass splitting in the stop sector, as given in  tab.~\ref{bestfit},
needs a very artificial fine tuning of the few free parameters of the CMSSM, especially for the
trilinear and Yukawa coupling in the stop sector, which can drive the diagonal elements of the stop
matrix, eq.~\ref{sqmatrix}, apart.
Note that the off-diagonal elements of this matrix are too small to generate
a large splitting, since
the left-handed stop is considerably heavier than the top,
implying that 
one of the diagonal elements  is considerably larger  than the off-diagonal elements.

In addition, problems arise with electroweak symmetry breaking,
since this requires $\mu > M_2$, while $R_b$ requires $\mu < M_2$ for a significant improvement.
In conclusion, within the CMSSM neither the high value of $R_b$ nor the low value of $R_c$ can be
explained, so in this scenario the solution must be sought in common systematic errors for all experiments, like PDB
branching ratios in the charm - sector~\cite{rberr} causing the too high value of $R_b$ and too low value of $R_c$.

\section{Conclusions}
The MSSM provides a good description of all electroweak data.
%and can simultanously give
%a prediction of $b\rightarrow s\gamma$ for high as well as for low $\tan\beta$ regions of
%the parameter space.
$R_b$ values up to 0.2193 are possible,
but the $b\rightarrow s\gamma$ rate is too low in this case.
With all mass bounds and the $b\rightarrow s\gamma$ rate included in the fit, the best
$\chi^2/d.o.f$ in the MSSM is  16.9/12, which  corresponds to a
probability of 15\% as  compared to the SM probability of about 8\%.
This best MSSM fit with light stops and light charginos is obtained for $\tan\beta=1.6$. Another
solution with $\tan\beta=50$ and light Higgses has a probability of 10\%, which is not much
of an improvement over the SM.

The enhancement of $R_b$ is compensated by   a decrease in $\alpha_s(M_Z)$ from 0.1215 in the SM case
to 0.116 in the MSSM. The latter is in somewhat better agreement with precise measurements
from DIS at low energies ($\alpha_s(M_Z)=0.112\pm0.005$).

The best solutions predict chargino, stop and pseudoscalar Higgs masses which may be detectable at LEP II.
The high $\tan\beta$ solution, requiring light Higgses, can certainly be excluded at LEP II, if no Higgses
are found. For the low $\tan\beta$ scenario
it will be more difficult to exclude
the SUSY explanation of the too high value of $R_b$, since even for chargino and stop masses above
95~GeV moderate improvements are still  possible.
On the other hand, it has to be pointed out that
the large splittings within the stop sector, which yield the best fit, are very difficult
to obtain in a natural way within a constrained MSSM model, since they require a very special
finetuning of the trilinear and Yukawa couplings.

\clearpage

\begin{table} [p]
\begin{center}
\protect\begin{sideways}
\begin{minipage}[b]{\textheight}
  \begin{tabular}{|l|c||cc||cc|cc|cc|}
\hline
\hline
Symbol & measurement & \multicolumn{8}{|c|}{ best fit observables}\\
\hline
\hline
 & & SM & & \multicolumn{6}{|c|}{MSSM}\\
\hline
$\tan\beta$ and pull        &                    &    & pull  & 1.0 & pull  & 1.6 & pull & 50 & pull\\
\hline
\hline
\hline
$M_Z$ [GeV]                 &  91.1884$\pm$ 0.0022 & 91.1882 &0.089&  91.1884  &0.000& 91.1884   &0.000 & 91.1884 &0.000\\
$\Gamma_{Z}$[GeV]           &  2.4964 $\pm$ 0.0032 & 2.4973  &-0.284&   2.4956 &0.241& 2.4958    &0.202 & 2.4952  &0.363\\
$\sigma_h$ [nb]             &  41.49  $\pm$ 0.078  & 41.452  &0.483&  41.441   &0.634&  41.448   &0.540 & 41.429  &0.777\\
$R_l$                       &  20.789 $\pm$ 0.032  & 20.774  &0.456&  20.7878  &0.037&  20.7878  &0.037 & 20.7898 &-0.024\\
$A_{FB}^l$                  &  0.0171 $\pm$ 0.0011 & 0.0164  &0.675&   0.0165  &0.577&   0.0164  &0.626 & 0.0165  &0.505\\
$R_b$                       &  0.2211 $\pm$ 0.0016 & 0.2159  &3.222&   0.2193  &1.128&   0.2180  &1.915 & 0.2180  &1.967\\
$R_c$                       &  0.1598 $\pm$ 0.0069 & 0.1723  &-1.805&   0.1703 &-1.522&   0.1706 &-1.559& 0.1706  &-1.565\\
$A_{FB}^b$                  &  0.1002 $\pm$ 0.0028 & 0.1034  &-1.145&   0.1040 &-1.366&   0.1038 &-1.286& 0.1043  &-1.453\\
$A_{FB}^c$                  &  0.0759 $\pm$ 0.0051 & 0.0740  &0.374&   0.0743  &0.319&   0.0741  &0.347 & 0.0745  &0.281\\
${\cal A}_b$                &  0.842  $\pm$ 0.052  & 0.9336  &-1.761&   0.9361 &-1.810&   0.9356 &-1.800& 0.9361  &-1.809\\
${\cal A}_c$                &  0.6180 $\pm$ 0.091  & 0.6680  &-0.550&   0.6684 &-0.554&   0.6682 &-0.552& 0.6685  &-0.555\\
\hline                                                                                           
${\cal A}_{\tau}$           &  0.1394 $\pm$ 0.0069 & 0.1477  &-1.196& 0.1482   & -1.270& 0.1479  &-1.235&0.1479 &-1.229\\
${\cal A}_e$                &  0.1429 $\pm$ 0.0079 & 0.1477  &-0.605& 0.1482   & -0.667 &0.1479  &-0.636&0.1484 &-0.712\\
$\sin^2\Theta_{eff}^{lept}(\langle Q_{FB}\rangle)$ &  0.2320  $\pm$ 0.0010& 0.2314 & 0.562 &0.23138  &0.624 & 0.23141 &0.593& 0.23133 &0.669\\
$M_W$ [GeV]                 &  80.33  $\pm$ 0.15   & 80.370  &-0.265& 80.417  &-0.583 &80.422   &-0.616 &80.452    &-0.814\\
%$\alpha_s(M_Z)$             &  0.117  $\pm$ 0.005  & 0.1215  &1.133& 0.1104  &3.913 &  0.1161   &2.465    &0.1162    &2.457\\
$1-M_W^2/M_Z^2$     &  0.2257 $\pm$ 0.0047 & 0.2232  &0.531& 0.2223  & 0.727 & 0.2222   &0.747    &0.2216    &0.869\\
$m_t$ [GeV]                 &  175    $\pm$ 9.     & 167.3   &0.858& 172.8   & 0.239 & 172.1    &0.32    &168.0     &0.776\\
$\sin^2\Theta_{eff}^{lept}(A_{LR})$(SLD) &  0.23049$\pm$ 0.0005 & 0.23144 &-1.900 & 0.23138 & -1.773 & 0.23141 &-1.834 & 0.23133 &-1.682\\
\hline                    
\end{tabular}             
\caption{\label{fitresults} Measurements of the observables \protect\cite{Lep2}
and the predicted results of the fits with minimum $\chi^2$. The pulls are
defined by (measurement - predicted value) / error of the measurement.}
\end{minipage}
\end{sideways}
\end{center}
\end{table}

\begin{table}[ht]
%\protect\vspace{-1.cm}
\begin{center}
\begin{tabular}{|c||r||r|r|r|}
\hline
\hline
\multicolumn{4}{|c|}{ Fitted SUSY parameters and masses}\\   
\hline
\hline
Symbol &\makebox[3.0cm]{\bf{$\tan\beta$=1.0}}&\makebox[3.0cm]{\bf{$\tan\beta$=1.6}}&\makebox[3.0cm]{\bf{$\tan\beta$=50}}\\
\hline
& no $b\rightarrow s\gamma$ & \multicolumn{2}{|c|}{$b\rightarrow s\gamma$ inc.}\\
\hline
\hline
\vspace{-0.09cm}
 $m_t$[GeV]                     & 173$\pm$7            &172$\pm$6            &168$\pm$6 \\
 $\alpha_s$                     & 0.1104$\pm$0.0043    &0.1161$\pm$0.0038    &0.1162$\pm$0.0039 \\
 $M_2$[GeV]                     & 25$\pm$8             &36$\pm$23            & - \\
 $\mu$[GeV]                     & 35$\pm$53            &42$\pm$9             &76$\pm$28 \\
 $m_{\tilde t_2}$[GeV]         & 48$\pm$5             &48$\pm$5             &53$\pm$40 \\
 $\phi_{mix}$                   &-0.163$\pm$0.115      &-0.203$\pm$0.091     &0.0021$\pm$0.0054 \\
 $m_{A}$[GeV]                   &    -                 &        -                    &50$\pm$5 \\
\hline                                        
\hline                                        
\multicolumn{4}{|c|}{ Particle Spectrum}\\    
\hline                                        
\hline                                        
% $m_{\tilde t_1}$[GeV]          &  1028     & 1034    &1011\\
 $m_{\tilde t_1}$[GeV]          &  \multicolumn{3}{c|}{$\approx 1$~TeV}\\
\cline{2-4} 
 $m_{\tilde t_2}$[GeV]          &  48       & 48      &53\\
\cline{2-4} 
 $m_{\tilde q}$[GeV]            & \multicolumn{3}{c|}{{1~TeV}}\\
 $m_{\tilde l} $[GeV]            & \multicolumn{3}{c|}{{0.5~TeV}}\\
\cline{2-4} 
\hline                                         
 $ m_{\tilde{\chi_1^\pm}}$[GeV]  & 91        & 106  &1504\\
 $ m_{\tilde{\chi_2^\pm}}$[GeV]  & 81        & 69   &76\\
\hline                                        
 $ m_{\tilde{\chi_1^0}}$[GeV]    & 15        & 21    &73\\
 $ m_{\tilde{\chi_2^0}}$[GeV]    & 35        & 38    &79\\
 $ m_{\tilde{\chi_3^0}}$[GeV]    & 90        & 97    &714\\
 $ m_{\tilde{\chi_4^0}}$[GeV]    & 102       & 102   &1504\\
\hline                                        
 $ m_{h}$[GeV]                   & 97        & 110    &50\\
\cline{2-3} 
 $ m_{H}$[GeV]                   &  \multicolumn{2}{c|}{{$\approx 1.5$~TeV}}   &98 \\
 $ m_{A}$[GeV]                   &  \multicolumn{2}{c|}{{$1.5$~TeV}}   &50 \\
 $ m_{H^\pm}$[GeV]               &  \multicolumn{2}{c|}{{$\approx 1.5$~TeV}}   &143\\
\cline{2-3} 
% $ m_{H}$[GeV]                   & 1506      & 1504    &98\\
% $ m_{A}$[GeV]                   & 1500      & 1500    &50\\
% $ m_{H^\pm}$[GeV]               & 1502      & 1502    &143\\
\hline                                        
 ${M_W^\pm}$[GeV]                & 80.4174   & 80.4224 &80.4520\\
 $\frac{BR(b\rightarrow s\gamma)}{BR(b\rightarrow ce\bar{\nu})}/10^{-4}$
                                 & (0.19) & 2.05  &2.30\\
\hline                                                       
\hline                                                       
 ${\chi^2}/d.o.f.$         & (15.1/11)        & 16.9/12   &18.4/12\\
\hline                                                       
 Probability               & 18\%             & 15\%      & 10\% \\
\hline                                                       
\hline                                                       
\end{tabular}
\end{center}
\caption[]{\label{bestfit}
Values of the fitted parameters (upper part) and
corresponding mass spectrum (lower part). The errors on the parameters
are parabolic ones. The parameters
given in the first column gave a minimum $\chi^2/d.o.f.$ of
$15.1/11$ for the LEP - observables. Here  $b\rightarrow s\gamma$
was not included in the fit, but the resulting $b\rightarrow s\gamma$ rate is about
one order of magnitude too small.
Including  $b\rightarrow s\gamma$ rate as measured by the CLEO
Collaboration \protect\cite{cleo} requires a higher value of $\tan\beta$, which
reduces the best $\chi^2/d.o.f.$ to $16.9/12$ (second column).
On the right hand side the results of the optimization for $\tan\beta=50$ are given.
The dashes indicate irrelevant parameters which were chosen high.
}    
\end{table}

\clearpage

\vspace{10cm}

\clearpage

%
%plots for the high-tan beta solution
%

\clearpage

\clearpage
\newpage
\section{References}                                                            
\begin{thebibliography}{20}
% 
%\bibitem{yel1}
%  Theoretical Physics and Particle Physics Experiments Division,\\
\bibitem{yel1} \it Proceedings of the Workshop Physics at LEP2 \rm,
               Editors G.~Altarelli, T.~Sj\"ostrand, F.~Zwirner,
               Vol.1 and Vol.2, CERN 96-01
\bibitem{boufi} M. Boulware, D. Finnell, \it Radiative Corrections to $BR (Z\rightarrow b\bar b)$ in the Minimal 
                Supersymmetric Standard Model\rm, \it Phys. Rev. \rm \bf D44\rm (1991) 2054
%\bibitem{chan1} P. H. Chankowski, S. Pokorski, \it Precision Tests of the MSSM\rm,\\
%                MPI-PhT/95-49, hep-ph 9505308
\bibitem{chan2} P. H. Chankowski, S. Pokorski, \it Chargino Mass and $R_b$ Anomaly \rm,\\
                IFT-96/6, hep-ph 9603310
\bibitem{ell1} J. Ellis, J. L. Lopez, D. V. Nanopoulos, hep-ph/9512288;
%  \it Phys. Lett. \rm  \bf  B286\rm  (1992) 85;
%               \it Nucl. Phys. \rm \bf B393\rm  (1993) 3;
%               \it Phys. Lett. \rm\bf  B324 \rm (1994) 173, \it Phys. Lett. \it \bf B333\rm  (1994) 118
\bibitem{garc3} D. Garcia, J. Sola,\it The Quantum Correlation $R_b-R_c$ in the MSSM: More Hints of Supersymmetry? \rm
                \it Mod. Phys. Lett. \rm \bf  A9\rm  (1994) 211
%\bibitem{lang1} P. Langacker, M. Luo, \it Phys.Rev. \rm \bf D44 \rm  (1991) 817
%\bibitem{alta1} G. Altarelli, R. Barbieri, F. Caravaglios, \it Nucl. Phys. \rm \bf B405\rm  (1993) 3,
%                \it Phys. Lett. \rm \bf  B314\rm  (1993) 357.
%\bibitem{wag1}  M. Carena, C.E.M. Wagner, CERN preprint CERN-TH-7393/94.
\bibitem{kan1} G.L. Kane, R.G. Stuart, J.D. Wells, \it A Global Fit of LEP/SLC Data with Light Superpartners\rm,
               \it Phys. Lett. \rm \bf B354\rm  (1995) 350,
                UM-TH-94-16, hep-ph/9505207
\bibitem{kan2} J.D. Wells, C. Kolda, G.L. Kane, \it Implications of $\Gamma(Z\rightarrow b\bar b)$ for Supersymmetry
                Searches and Model-Building \rm, \it Phys. Lett. \rm \bf B338\rm (1993) 219,
                UM-TH-94-23, hep-ph/9408228
\bibitem{garcia} D. Garcia, R. Jimenez, J. Sola,
                 \it Supersymmetric Electroweak Renormalization of the Z Width in the MSSM.\rm
                 \it Phys. Lett. \rm \bf B347 \rm (1995) 309;
                 \it Phys. Lett. \rm \bf B347 \rm (1995) 321
\bibitem{garcia2} D. Garcia, J. Sola, \it Matching the low-Energy and the high-Energy Determinations of
                 $\alpha_s$(M(Z)) in the MSSM,\rm \it  Phys. Lett. \rm \bf B357\rm  (1995) 349
\bibitem{minuit}
                F. James, \it MINUIT Reference Manual\rm, Version~94.1,
                Computing and Networks Division
                CERN Geneva, Switzerland  
\bibitem{tevatron} F. Abe et al., \it CDF Collaboration, \rm Measurements of the
               W Boson Mass, \it Phys. Rev. \rm \bf D52 \rm  (1995) 4784;\\
%               FERMILAB-PUB-95/033-E and FERMILAB-PUB-95/035-E-1995. \\
               C. K. Jung, \it D\O \  Collaboration, \rm W Mass Measurements from D\O \ and
               CDF Experiments at the Tevatron, talk given at the 27th ICHEP, Glasgow,
               Scotland, 20-27 July 1994.
\bibitem{top} F. Abe et al., \it CDF Collaboration, \it Phys. Rev. Lett. \rm \bf 74 \rm (1995) 2626,
%  \rm FERMILAB-PUB-95-022-E,
              March 1995.  \\
              S. Abachi et al., \it D\O \ Collaboration,  \it Phys. Rev. Lett. \rm \bf 74 \rm (1995) 2632,
%              \rm FERMILAB-PUB-95-028-E,
              March 1995.;\\
%              D0 Collaboration, preliminary result presented by U.Heintz, Les Rencotres
%              de Physique de la Vallee D'Aoste, La Thuile, March 1996\\
              CDF Collaboration, A.Caner, presented at Les Rencontres de Physique de la Vallee
              d'Aoste, La Thuile, March 1996\\
              D0 Collaboration, M.Narain, presented at Les Rencontres de Physique de La Vallee
              d'Aoste, La Thuile, March 1996
\bibitem{Lep1} LEP Electroweak Working Group, \it CERN-preprint \rm LEPEWWG/95-01,
               March 1995.
\bibitem{Lep2} LEP Electroweak Working Group, \it CERN-preprint \rm LEPEWWG/96-01,
               March 1996.
\bibitem{cleo} CLEO-Collaboration, R.~Ammar et al., {\it Phys. Rev. Lett.} {\bf 74},
               {\it (1995) 2885}
\bibitem{lim1}  ALEPH Collaboration, \it Search for supersymmetric particles on $e^+e^-$
                collisions at centre-of-mass energies of 130 and 136~GeV\rm,
                CERN-PPE/96-10
\bibitem{lim2}  OPAL Collaboration, \it Topological Search for the Production of Neutralinos
                and Scalar Particles\rm, CERN-PPE/96-019;\\
                \it Search for Chargino and Neutralino Production Using the OPAL Detector
                at $\sqrt{s}=$ 130 - 136~GeV\rm, CERN-PPE/96-020;\\
               DELPHI Collaboration, \it Search for the Lightest Chargino at $\sqrt{s} =$ 130 and 136~GeV \rm,
               CERN-PPE/96-75
\bibitem{lim4} D0 Collaboration, \it Search for Light Top Squarks in $p\bar p$ Collisions at 1.8 TeV\rm,
                    \it Phys. Rev. Letters \rm  \bf 76\rm, 2222 (1996) ,
               FERMILAB-PUB-95/380-E;\\
               DELPHI Collaboration, \it Search for neutralinos, scalar leptons and scalar quarks
               in $e^+e^-$ interactions at $\sqrt s=$130~GeV and 136~GeV\rm,
               in preparation
%\bibitem{lim5}  DELPHI Collaboration, \it Search for the Standard Model Higgs boson in $Z^0$ decays\rm,
%                CERN PPE/94-46/Rev.
\bibitem{lim6}  ALEPH Collaboration, \it Mass Limit for the Standard Model Higgs Boson
                with the full LEP I ALEPH Data Sample\rm , CERN PPE/96-079
\bibitem{rev96} \it Particles and Fields\rm,
                \it Phys. Rev. \rm \bf D50 \rm (1994), 1173-1826, Number 3;\\
                R.M. Barnett et al., \it Phys. Rev. \rm \bf D54 \rm (1996) 1.
\bibitem{sirlin} A. Sirlin, \it Phys. Rev. \rm \bf D 22 \rm (1980) 971.  \\
                 W. J. Marciano and A. Sirlin, \it Phys. Rev. \rm \bf D 22 \rm (1980)
                 2695.
\bibitem{sola} D. Garcia and J. Sol\`{a}, \it Mod. Phys. Lett. \rm \bf A 9 \rm
               (1994) 211. \\
               P.H. Chankowski, A. Dabelstein, W. Hollik, W. M\"osle,
               S. Pokorski and J. Rosiek, \it Nucl. Phys. \rm \bf B 417 \rm 
               (1994) 101.
\bibitem{bhp} For a recent review see: \it Precision Calculations for the $Z$ Resonance,
              \rm Yellow report CERN 95-03, eds. D. Bardin, W. Hollik and G. Passarino,
              and references therein.
\bibitem{fleischer} L. Avdeev, J. Fleischer, S. Mikhailov and O. V. Tarasov, \it
                    Phys. Lett. \rm \bf B 336 \rm (1994) 560. \\
                    J. Fleischer, O. V. Tarasov and F. Jegerlehner, \it Phys. Lett.
                    \rm \bf B 319 \rm (1993) 249;\\
                    R. Barbieri, M. Beccaria, P. Ciafaloni, G. Curci. A. Vicere,
                    \it Phys. Lett. \rm\bf B288 \rm (1992) 95; \it Nucl. Phys. \rm \bf  B409 \rm (1993) 105;\\
                    K.G. Chetyrkin, J.H. Kuehn, M. Steinhauser,
                    \it Phys. Lett. \rm \bf B351 \rm (1995) 331;\\
                    J. Fleischer, F. Jegerlehner, P. Raczka, O.V. Tarasov,
                    \it Phys. Lett. \rm \bf B293 \rm (1992) 437;\\
                    G. Buchalla, A.J. Buras, \it Nucl. Phys. \rm \bf  B398 \rm (1993) 285
\bibitem{dab/hollik} A. Dabelstein, W. Hollik, W. M\"osle, in preparation.
\bibitem{Chet/Kw} K. G. Chetyrkin, J. H. K\"uhn and A. Kwiatkowski, \it Phys. Lett. 
                  \rm \bf B282 \rm (1992) 221; \\
                  K. G. Chetyrkin and A. Kwiatkowski, \it Phys. Lett. \rm \bf B305
                  \rm (1993) 285; \\
                  K. G. Chetyrkin, A. Kwiatkowski and M. Steinhauser,
                  \it Mod. Phys. Lett. \rm \bf A 29 \rm (1993) 2785; \\
                  A. Kwiatkowski, M. Steinhauser, \it Phys.~Lett.\
                  \rm \bf B344 \rm (1995) 359; \\ 
                  K. G. Chetyrkin, J. H. K\"uhn and A. Kwiatkowski,
                  in: \it Precision Calculations for the Z Resonance,
                  \rm CERN 95-03, eds.\ D. Bardin, W. Hollik,
                  G. Passarino\\
                  S. Peris, A. Santamaria,  CERN-TH-95-21 (1995).
\bibitem{bhs} M. B\"ohm, W. Hollik and H. Spiesberger, \it Fortschr. Phys.
              \rm \bf 34 \rm (1986) 687. \\
              W. Hollik, \it Fortschr. Phys. \rm \bf 38 \rm (1990) 165.
\bibitem{dghk} A. Denner, R. Guth, W. Hollik, J.H. K\"uhn,
               \it Z.~Phys.\  \rm \bf C51 \rm (1991)  695
\bibitem{dabelstein} A. Dabelstein, \it Z.~Phys. \rm \bf C67  \rm  (1995) 495;
                                    \it Nucl.~Phys. \rm\bf  B456 \rm  (1995) 25
\bibitem{hunter}  H. P. Nilles, \it Phys. Rep. \rm \bf 110  \rm (1984) 1. \\
      H. E. Haber and G. Kane, \it Phys. Rep. \rm \bf 117  \rm (1985) 75.\\
      J. F. Gunion and H. E. Haber, \it Nucl. Phys. \rm \bf B272  \rm (1986) 1;
      \it Nucl. Phys. \rm \bf B402 \rm (1993) 567. \\
      J. F. Gunion, H. E. Haber, G. Kane and S. Dawson: \it
      The Higgs Hunter's Guide, \rm  Addison-Wesley 1990.
\bibitem{ellisetal} J. Ellis, G. Ridolfi and F. Zwirner, \it Phys. Lett.
                 \rm \bf B257 \rm
               (1991) 83. 
\bibitem{zwid1} R. Barbieri, G. Gamberini, G. Giudice, G.Ridolfi,
                \it Signals of Supersymmetry at the $Z_0$ Resonance \rm
                \it Nucl. Phys. \rm \bf  B296 \rm (1988) 75-90.
\bibitem{bsgamma}
              R. Barbieri and G. Giudice, {\it Phys. Lett.} {\bf B309} (1993) 86;\\
              R. Garisto and J.N. Ng, {\it Phys. Lett.} {\bf B315} (1993) 372\rm;\\
              S. Bertolini, F. Borzumati, A.Masiero, and G. Ridolfi,
              \it  Nucl. Phys. \rm  {\bf B353} (1991) 591 {\em and references therein};\\
               N. Oshimo, \it  Nucl. Phys. \rm {\bf B404} (1993) 20;\\
              S. Bertolini, F. Vissani,  \it Z. Phys. \rm \bf C67 \rm (1995) 513, 1995
\bibitem{buras} A. J. Buras et al., \it Nucl. Phys. \rm \bf B424\rm (1994) 374
\bibitem{zfitter} D. Bardin et al., \it ZFITTER, An Analytical Program for Fermion Pair Production in $e^+e^-$ Annihilation\rm,
                CERN-TH.6443/92
\bibitem{ralf}  R. Ehret, \it Die Bestimmung der Kopplungskonstanten $\alpha_s$ am LEP-Speicherring
                und Tests von gro\ss en Vereinigungstheorien \rm,  Ph.D. Thesis,
                 IEKP-KA/95-13\rm
%\bibitem{renton} Peter B. Renton, \it Review of experimental results on precision tests of electroweak
%                theories \rm ,  CERN-PPE/96-63 
\bibitem{cmssm} W.~de~Boer et al., \it Combined Fit of Low Energy Constraints to Minimal Supersymmetry
                and Discovery Potential at LEP II\rm, hep-ph/9603350
\bibitem{wimhig} W.~de~Boer et al., \it MSSM predictions of the Neutral Higgs Boson Masses And LEP- II
                 Production Cross- Sections \rm, hep-ph/9603346 and references therein
\bibitem{lattbb} B. Grinstein, I.Z. Rothstein \it Errors in Lattice Extractions of $\alpha_s$
                 due to Use of Unphysical Pion Masses, \rm UCSD-TH-96-09,
                 hep-ph/9605260
\bibitem{rberr} I. Dunietz, J. Incandela, F.D. Snider, K.Tesima, I. Watanabe,
                \it Comments on Recent Measurements of $R_c$ and $R_b$ \rm, 
                FERMILAB-PUB-96/26-T, hep-ph/9606327\\
                P. Paganini, P. Roudeau, A. Stocchi ,
                \it An heretic evaluation of the accuracy on R$_b$ \rm,
                DELPHI note 96-2 PHYS 586\\
                P. Paganini, P. Roudeau and A. Stocchi ,
                \it What is really interesting in the measurement of $R_{c}$ ? \rm,
                DELPHI note 96-66 PHYS 626 
%
%\bibitem{Beenakker} A. A. Akhundov, D. Bardin and T. Riemann, \it Nucl. Phys. \rm
%             \bf B 276 \rm (1986) 1. \\
%             W. Beenakker and W. Hollik, \it Z. Phys. \rm \bf C 40 \rm (1988) 141. \\
%             J. Bernabeu, A. Pich and A. Santamaria, \it Phys. Lett. \rm \bf B 200 \rm
%             (1988) 569.  
%\bibitem{ku/kni}  J. H. K\"uhn and B. A. Kniehl, \it Phys. Lett. \rm \bf B 224 \rm
%                  (1990) 229; \it Nucl. Phys. \rm \bf B 329 \rm (1990) 547.
%\bibitem{tarasow} J. Fleischer, O. V. Tarasow, F. Jegerlehner and P. R\c{a}czka, \it
%               Phys. Lett. \rm \bf B 293 \rm (1992) 437. \\
%               G. Buchalla and A. J. Buras, \it Nucl. Phys. \rm \bf B 398 \rm
%               (1993) 285. \\
%               G. Degrassi, \it Nucl. Phys. \rm \bf B 407 \rm (1993) 271.
%\bibitem{barbieri} R. Barbieri, M. Beccaria, P. Ciafaloni, G. Curci and A. Vicere,
%              \it Phys. Lett. \rm \bf B 288 \rm (1992) 95; \it Phys. Lett. \rm \bf 
%              B 312 \rm (1993) 511. \\
%              A. Denner, W. Hollik and B. Lampe, \it Z. Phys. \rm \bf C 60 \rm 
%              (1993) 193.
%              R. Barbieri, P. Ciafaloni and A. Strumia, \it Phys. Lett. \rm \bf 
%              B 317 \rm (1993) 381.
%\bibitem{hara} L. Hall and L. Randall, \it Phys. Rev. Lett. \rm \bf 65 \rm
%               (1990) 2939. 
%\bibitem{boer} J. Ellis, S. Kelly and D. Nanopoulos, \it Phys. Lett. \rm \bf B 249
%               \rm (1990) 441. \\
%               U. Amaldi, W. de. Boer and H. F\"urstenau, \it Phys. Lett. \rm \bf
%               B 260 \rm (1991) 447. \\
%               P. Langacker and M. Luo, \it Phys. Rev. \rm \bf D 44 \rm (1991) 477.
%\bibitem{Denner} A. Denner, R. J. Guth, W. Hollik and J. H. K\"uhn,
%                 \it  Z. Phys. \rm \bf C51 \rm (1991) 695.
%\bibitem{Cornet}  F. Cornet, W. Hollik, W. M\"osle,
%                  \it Nucl. Phys. \rm \bf B 428 \rm (1994) 61.
%\bibitem{Boulware} M. Boulware and D. Finnell, \it Phys. Rev. \rm \bf D 44 \rm 
%                   (1991) 2054. \\
%                  B. Q. Hu, J. M. Yang and C. S. Li, \it Commun. Theor. Phys.
%                 \rm \bf 20 \rm (1993) 213.
%\bibitem{Wells} J. D. Wells, C. Kolda and G. L. Kane,
%                \it Phys. Lett. \rm \bf B 338 \rm (1994) 219.
%\bibitem{Garcia1} D. Garcia, R. A. Jim\'enez and J. Sol\`a, 
%                  \it preprint \rm UAB-FT-343, Sep. 1994; 
%                 \it Phys.Lett \rm \bf B347\rm (1995) 309-320,
%                 UAB-FT-344, Sep. 1994;
%                 \it Phys.Lett \rm \bf B347\rm (1995) 321-331,
%                 UAB-FT-358, Jan. 1995.
%                 \it Phys.Lett \rm \bf B354\rm (1995) 335-344

%\bibitem{carena} M. Carena and C. E. M. Wagner, \it CERN-preprint \rm CERN-TH-7393-94,
%                 Jul. 1994.
%\bibitem{Bhatta} L. Chongsheng and H. Bingquan, \it Chinese Phys. Lett. \rm \bf 
%                 6 \rm (1992) 289. \\
%                 G. Bhattacharyya and A. Raychaudhuri, \it Phys. Rev. \rm \bf D 47 
%                 \rm (1993) 2014. \\
%                 A. Djouadi, M. Drees and H. K\"onig, \it Phys. Rev. \rm \bf D 48 
%                 \rm (1993) 3081. \\
%                 C. S. Li, B. Q. Hu, J. M. Yang and Z. Y. Fang, \it J. Phys. \rm 
%                 \bf G 19 \rm (1993) 13. \\
%                 L. Clavelli, \it Alabama-preprint \rm UAHEP-948, Oct. 1994.
%\bibitem{QCD}  S. Bethke, \it Proceedings of the Linear Collider Workshop, \rm
%                 Waikoloa 1993. \\
%               S. Catani, \it Proceedings of the EPS Conference, Marseille 1993, \rm eds.
%               J. Carr and M. Perrottet. 
%\bibitem{hempf} H. E. Haber and R. Hempfling, \it Phys. Rev. Lett. \rm \bf 66 \rm
%               (1991) 1815. \\
%                R. Hempfling and A. H. Hoang, \it Phys. Lett. \rm \bf B 331 \rm
%                (1994) 99.  \\
%                J.A. Casas, J.R. Espinosa, M. Quiros and A. Riotto, \it CERN-preprint 
%                \rm CERN-TH. 7334/94 (July 1994).
%\bibitem{djouadi/gl} A. Djouadi and M. Drees, \it Madison-preprint \rm MAD-PH-84-853,
%                    Oct. 1994. 
%\bibitem{altarelli} G. Altarelli, R. Barbieri and F. Caravaglios, \it Nucl. Phys. \rm
%                    \bf B 405 \rm (1993) 3.
%\bibitem{Djouadi} A. Djouadi, G. Girardi, C. Verzegnassi, W. Hollik and
%                  F. M. Renard, \it Nucl. Phys. \rm \bf B 349 \rm (1991) 48.
%\bibitem{Hollik1} W. Hollik, 
%                  \it Precision Tests of the Standard Model, \rm ed. P. Langacker
%             (Advanced Series on Directions in High Energy Physics, World Scientific
%             Publishing Co.).
%
\end{thebibliography}
\end{document}